\documentclass[aps,superscriptaddress,twocolumn,twoside,floatfix,pra,nofootinbib,a4paper]{revtex4}
\usepackage{times}
\usepackage{epsfig}
\usepackage{amsfonts}
\usepackage{amsmath}
\usepackage{amssymb}
\usepackage{amsthm}
\usepackage{color}
\usepackage{multirow}
\usepackage[normalem]{ulem}
\newcommand{\stkout}[1]{\ifmmode\text{\sout{\ensuremath{#1}}}\else\sout{#1}\fi}
\usepackage{latexsym}
\usepackage{mathrsfs}	
\usepackage{natbib}
\usepackage{verbatim}
\usepackage[T1]{fontenc}
\usepackage{float}

\usepackage{enumitem}

\usepackage{graphicx}
\usepackage{xcolor}

\usepackage[colorlinks=true,linkcolor=blue,citecolor=magenta,urlcolor=blue]{hyperref}

\newcommand{\ket}[1]{|#1\rangle}

\newcommand{\ketbra}[2]{|#1\rangle\langle#2|}

\newcommand{\beq}{\begin{equation}}
\newcommand{\eeq}{\end{equation}}
\newcommand{\beqa}{\begin{eqnarray}}
\newcommand{\eeqa}{\end{eqnarray}}

\begin{document}
	

\title{Bilocal Bell inequalities violated by the quantum Elegant Joint Measurement}


\author{Armin Tavakoli}
\affiliation{D\'epartement de Physique Appliqu\'ee, Universit\'e de Gen\`eve, CH-1211 Gen\`eve, Switzerland}

\author{Nicolas Gisin}
\affiliation{D\'epartement de Physique Appliqu\'ee, Universit\'e de Gen\`eve, CH-1211 Gen\`eve, Switzerland}

\author{Cyril Branciard}
\affiliation{Universit\'e Grenoble Alpes, CNRS, Grenoble INP, Institut N\'eel, 38000 Grenoble, France}

\begin{abstract}
Network Bell experiments give rise to a form of quantum nonlocality that conceptually goes beyond Bell's theorem. We investigate here the simplest network, known as the bilocality scenario. We depart from the typical use of the Bell State Measurement in the network central node and instead introduce a family of symmetric iso-entangled measurement bases that generalise the so-called Elegant Joint Measurement. This leads us to report noise-tolerant quantum correlations that elude bilocal variable models. Inspired by these quantum correlations, we introduce network Bell inequalities for the bilocality scenario and show that they admit noise-tolerant quantum violations. In contrast to many previous studies of network Bell inequalities, neither our inequalities nor their quantum violations are based on standard Bell inequalities and standard quantum nonlocality. Moreover, we pave the way for an experimental realisation by presenting a simple two-qubit quantum circuit for the implementation of the Elegant Joint Measurement and our generalisation. 
\end{abstract}
	
	
	\maketitle
	

\textit{Introduction.---} The violation of Bell inequalities is a hallmark property of quantum theory. It asserts that the predictions of quantum theory cannot be accounted for by any physical model based only on local variables~\cite{Bell}. Such violations, referred to as quantum nonlocality, do not only provide insights in the foundations of quantum theory, but they also constitute a powerhouse for a broad scope of applications in quantum information science~\cite{Review}.
	
A standard Bell experiment features a source that emits a pair of particles shared between two space-like separated observers who perform local and independent measurements. In quantum theory the particles can be entangled, thus enabling  global randomness~\cite{GisinQchance14}. In contrast, in local variable  models aiming to simulate the quantum predictions, the particles are endowed with classically correlated  stochastic properties  that locally determine the outcome of a given measurement. Many decades of research on Bell inequalities have brought a relatively deep understanding of quantum nonlocality and have established standard methods for characterising correlations in both quantum models and local variable models~\cite{Review}.

The last decade witnessed a significant conceptual advance: much attention was directed at going beyond correlations in standard Bell experiments in favour of investigating correlations in networks featuring many observers and several independent sources of particles~\cite{Branciard, Fritz}. While a standard Bell experiment may be viewed as a trivial network (with a single source), the introduction of multiple independent sources is conceptually interesting  since it brings into play new physical ingredients and corresponds to the topology of future quantum networks. In contrast to standard Bell experiments, network Bell experiments feature some observers who hold independent particles (from different sources) and therefore \emph{a priori} share no correlations. Moreover, entanglement can be distributed in the network, in particular to initially independent observers, through the process of entanglement swapping~\cite{Zukowski}. Recent years have seen much attention being directed at characterising classical, quantum and post-quantum correlations in networks, many times through the construction of network Bell inequalities and the exploration of their quantum violations~\cite{Branciard2, ChavesFritz, Tavakoli, Henson, TavakoliConnected, Rosset, Chaves, Tavakoli2, Andreoli, Tavakoli3, Fraser, Luo, Inflation, Salman, Wolfe, GisinBancal, Fritz2, Wood, Renou, ChavesKueng}. In general, this is challenging due to the fact that the introduction of multiple independent sources makes the set of local variable correlations non-convex~\cite{Branciard}.
	
\begin{figure}
		\includegraphics[width=\columnwidth]{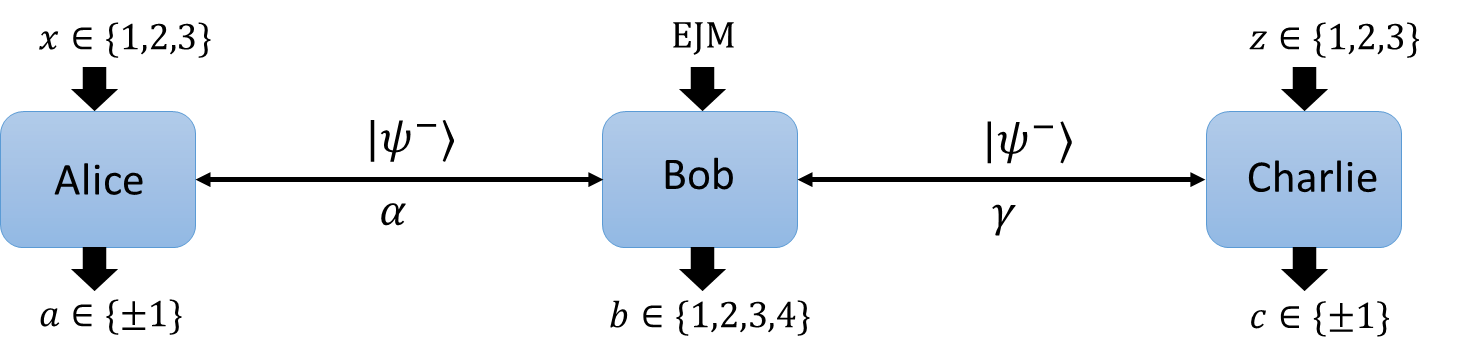}
		\caption{Bilocality scenario: Bob independently shares a ``state'' with Alice and Charlie, respectively.  In a quantum experiment, these are independent, typically entangled quantum states ($\ket{\psi^-}$), while in a bilocal model these are associated to independent local variables ($\alpha$ and $\gamma$).}\label{FigScenario}
\end{figure}

Here, we focus on the simplest nontrivial network Bell experiment, known as the \textit{bilocality scenario}. It features two independent sources that each produce a pair of particles. The first pair is shared between observers Alice and Bob while the second pair is shared between Bob and another observer, Charlie (see Figure~\ref{FigScenario}). Interestingly, there are known Bell inequalities for the bilocality scenario (bilocal inequalities), i.e.~inequalities for the observed correlations that are satisfied by all local variable models respecting the independence of the two sources. Importantly, these inequalities are also known to admit quantum violations. The quantum violations typically arise from Bob implementing a Bell State Measurement (BSM, encountered in quantum teleportation~\cite{Teleport} and entanglement swapping~\cite{Zukowski}). Conspicuously, both the inequalities and their reported violations strongly resemble those encountered in the standard Bell experiments (see e.g.~\cite{Branciard2, Gisin, Tavakoli2, Andreoli}). For instance the standard bilocal inequality, first presented in Ref.~\cite{Branciard2}, is essentially built on the Clauser-Horne-Shimony-Holt (CHSH) inequality~\cite{CHSH} and its quantum violations through the BSM turn out to effectively correspond to Bob in a coordinated manner separately testing the CHSH inequality with Alice and Charlie respectively. Indeed, the BSM measurement amounts to performing simultaneously the two commuting measurements of $\sigma_1\otimes\sigma_1$ and $\sigma_3\otimes\sigma_3$ (where $(\sigma_1,\sigma_2,\sigma_3)$ are the three Pauli observables) on Bob's two independent qubits, and ample numerical evidence shows that the optimal measurements settings for Alice and Charlie are at $\pm45$ degrees on the Bloch sphere, i.e.~exactly those settings tailored for the CHSH inequality. Given this close resemblance to the CHSH inequality, it is perhaps unsurprising that the critical singlet visibility, required for two identical noisy singlet states to enable a violation, is the same as that encountered in the CHSH inequality, namely $\frac{1}{\sqrt{2}}$ for each state.

Here we investigate quantum nonlocality in the bilocality scenario that is not based on the BSM and does not directly trace back to standard quantum nonlocality as in the previous cases. To this end, we present a family of two-qubit entangled measurements generalising the so-called Elegant Joint Measurement (EJM)~\cite{EJM}. These allow Bob to effectively distribute (in an entanglement swapping scenario) different entangled states to Alice and Charlie from those obtained through a BSM. We investigate bilocal models for the resulting correlations, show explicit quantum violations of bilocality and obtain the critical visibility per singlet for a quantum violation. Subsequently, we introduce new bilocal inequalities tailored to our quantum correlations and show that they can detect quantum nonlocality in the network at reasonable singlet visibilities. Furthermore, towards experimental demonstrations of quantum violations of network Bell inequalities, that are not based on standard Bell inequalities, we explore the implementation of our generalised EJM. We prove that it cannot be implemented in linear optical schemes without auxiliary photons but that it can be implemented with a simple two-qubit quantum circuit.

\textit{Entangled measurements with tetrahedral symmetry.---} We consider symmetric entangled measurements on two qubits that, most naturally, have four outcomes. Specifically, we present a family of bases $\{\ket{\Phi_b^\theta}\}_{b=1}^4$ of the two-qubit Hilbert space, parametrised by  $\theta\in[0,\frac{\pi}{2}]$, such that all elements are equally entangled and, moreover, the four local states, corresponding to either qubit being traced out, form a shrunk regular tetrahedron inside the Bloch sphere.

To construct such bases, let us first introduce the pure qubit states $\ket{\vec{m}_b}$ that point (on the Bloch sphere) towards the four vertices
\begin{align}\label{tetra}\nonumber
& \vec{m}_1=\left(+1,+1,+1\right), \quad \vec{m}_2=\left(+1,-1,-1\right),\\
& \vec{m}_3=\left(-1,+1,-1\right), \quad \vec{m}_4=\left(-1,-1,+1\right)
\end{align}
of a regular tetrahedron, as well as the states $\ket{{-}\vec{m}_b}$ with the antipodal direction. Specifically, we write these tetrahedron vertices in cylindrical coordinates as $\vec{m}_b=\sqrt{3}\left(\sqrt{1-\eta_b^2}\cos\varphi_b, \sqrt{1-\eta_b^2}\sin\varphi_b,\eta_b\right)$ and define
\begin{equation}
\ket{{\pm}\vec{m}_b}=\sqrt{\frac{1\pm\eta_b}{2}}e^{-i\varphi_b/2}\ket{0}\pm\sqrt{\frac{1\mp\eta_b}{2}}e^{i\varphi_b/2}\ket{1}.
\end{equation}
Our family of generalised EJM bases, with the above properties, is then given by
\begin{equation}\label{Phi_b}
\ket{\Phi_b^\theta}=\frac{\sqrt{3}+e^{i\theta}}{2\sqrt{2}}\ket{\vec{m}_b,{-}\vec{m}_b}+\frac{\sqrt{3}-e^{i\theta}}{2\sqrt{2}}\ket{{-}\vec{m}_b,\vec{m}_b}
\end{equation}
Notice that for $\theta=0$, we obtain the EJM introduced in Ref.~\cite{EJM} (the largest local tetrahedron in our family, of radius $\frac{\sqrt{3}}{2}$), while for $\theta=\frac{\pi}{2}$, we obtain the BSM (the smallest local tetrahedron, of radius zero) up to local unitaries (which can for instance be chosen as $U_1 \otimes U_2=\openone \otimes e^{\frac{2\pi i}{3}\frac{\sigma_1+\sigma_2+\sigma_3}{\sqrt{3}}}$ to recover the standard BSM). By varying $\theta$, we thus  continuously interpolate between the EJM and the BSM.

\textit{Quantum correlations.---} We consider a specific quantum implementation of the bilocality experiment illustrated in Figure~\ref{FigScenario}. We let Bob apply the generalised EJM and consider that both sources emit pairs of  qubits corresponding to noisy singlets (so-called Werner states~\cite{werner})
\begin{equation}
	\rho_i=V_i\ketbra{\psi^-}{\psi^-}+\frac{1-V_i}{4}\openone,
\end{equation}
for $i\in\{1,2\}$ where $V_i\in[0,1]$ denotes the visibility of each singlet $\ket{\psi^-}=\frac{1}{\sqrt{2}}\left(\ket{0,1}-\ket{1,0}\right)$.  By applying his measurement onto distributed (pure) singlets, Bob effectively prepares Alice's and Charlie's joint state in an entangled state similar to that of Eq.~\eqref{Phi_b}, up to a change in signs for $\vec{m}_b$ and $\theta$. Due to the tetrahedral structure of the distributed states we expect to find strong correlations between Alice and Charlie when they perform  measurements of the three Pauli observables \cite{CommentNote}. We therefore let each of them have three possible measurement settings $x,z\in\{1,2,3\}$ (corresponding to the observables  $(\sigma_x, \sigma_z)$), with binary outcomes denoted $a,c\in\{+1,-1\}$.

To reflect the symmetry of our scenario, it is convenient to identify Bob's outcome $b$ with the corresponding vector $\vec m_b$ from Eq.~\eqref{tetra}, i.e., to write $b$ as $\pm 1$-valued 3-vector $b = (b^1, b^2, b^3)$.
The conditional probability distribution $p(a,b,c|x,z)$ obtained in the experiment can then be characterised in terms of the single-, two- and three-party correlators $\langle A_x\rangle$, $\langle B^y\rangle$, $\langle C_z\rangle$, $\langle A_xB^y\rangle$, $\langle B^yC_z\rangle$, $\langle A_xC_z\rangle$ ($=\langle A_x\rangle \langle C_z\rangle$ in the bilocality scenario) and $\langle A_xB^yC_z\rangle$ for all $x,y,z\in\{1,2,3\}$, with e.g. $\langle A_xB^yC_z\rangle = \sum_{a,b^1,b^2,b^3,c=\pm1} a \, b^y \, c \, p(a,b,c|x,z)$ and similarly for the other correlators~\cite{footnote_p_from_correlators}.
For the quantum correlation $p_\text{Q}^\theta$ obtained from our above choice of states and measurements, these correlators become
\begin{align}\label{correlators}\nonumber
& \langle A_x\rangle = \langle B^y\rangle = \langle C_z\rangle = \langle A_xC_z\rangle = 0, \nonumber \\[1mm]
& \langle A_xB^y\rangle=-{\textstyle \frac{V_1}{2}}\cos\theta \, \delta_{x,y}, \quad \langle B^yC_z\rangle={\textstyle \frac{V_2}{2}}\cos\theta \, \delta_{y,z}, \nonumber\\[1mm]
& \langle A_xB^yC_z\rangle=\begin{cases}
 -\frac{V_1V_2}{2}\left(1{+}\sin\theta\right) & \!\!\text{if } xyz\in\{123,231,312\}\\
 - \frac{V_1V_2}{2}\left(1{-}\sin\theta\right) & \!\!\text{if } xyz\in\{132,213,321\}\\
  \,0 & \!\!\text{otherwise}
\end{cases}\!,
\end{align}
where $\delta$ is the Kronecker symbol.

\textit{Simulating quantum correlations in bilocal models.---} Let us first investigate whether the quantum probability distribution $p_\text{Q}^\theta$ admits a bilocal model. In such a model, each pair of particles is associated to a local variable denoted $\alpha$ and $\gamma$ respectively (see Figure~\ref{FigScenario}). Alice's (Charlie's) outcome is determined by her (his) setting and $\alpha$ ($\gamma$). Since they each have three possible settings, we can without loss of generality represent the local variables as triples $\alpha = (\alpha_1,\alpha_2,\alpha_3)$ and $\gamma = (\gamma_1,\gamma_2,\gamma_3)$ with entries $\pm 1$, with each $\alpha_x, \gamma_z$ denoting Alice's or Charlie's deterministic outcome for the setting $x$ or $z$.
A bilocal model can thus be written as
\begin{equation}\label{biloc}
p_{\text{biloc}}(a,b,c|x,z)=\sum_{\alpha,\gamma} q_\alpha^{(1)}q_\gamma^{(2)}\delta_{a,\alpha_x}\delta_{c,\gamma_z}p(b|\alpha,\gamma),
\end{equation}
where $\{q^{(1)}_\alpha\}_\alpha$ and $\{q^{(2)}_\gamma\}_\gamma$ are probability distributions representing the stochastic nature of the local variables $\alpha$ and $\gamma$ respectively, and $p(b|\alpha,\gamma)$ are probability distributions representing the stochastic response of Bob upon receiving $(\alpha,\gamma)$.

The central question is whether the quantum correlations characterised by Eq.~\eqref{correlators} can be simulated in a bilocal model. We investigate the matter with three different approaches. Firstly, we set $V\equiv V_1=V_2$ (equal noise on both sources), and $\theta = 0$ (as in the original EJM~\cite{EJM}). By employing semidefinite relaxations of the set of bilocal correlations, one can obtain a necessary condition for the existence of a bilocal model~\cite{Pozas}. An evaluation of the relevant semidefinite program guarantees that a violation of bilocality is obtained whenever $V \gtrsim  83\,\%$~\cite{Pozas2}. However, this bound is not expected to be tight due to the non-convex nature of the set of quantum correlations with independent sources.

\begin{figure}[t!]
	\centering
	\includegraphics[width=\columnwidth]{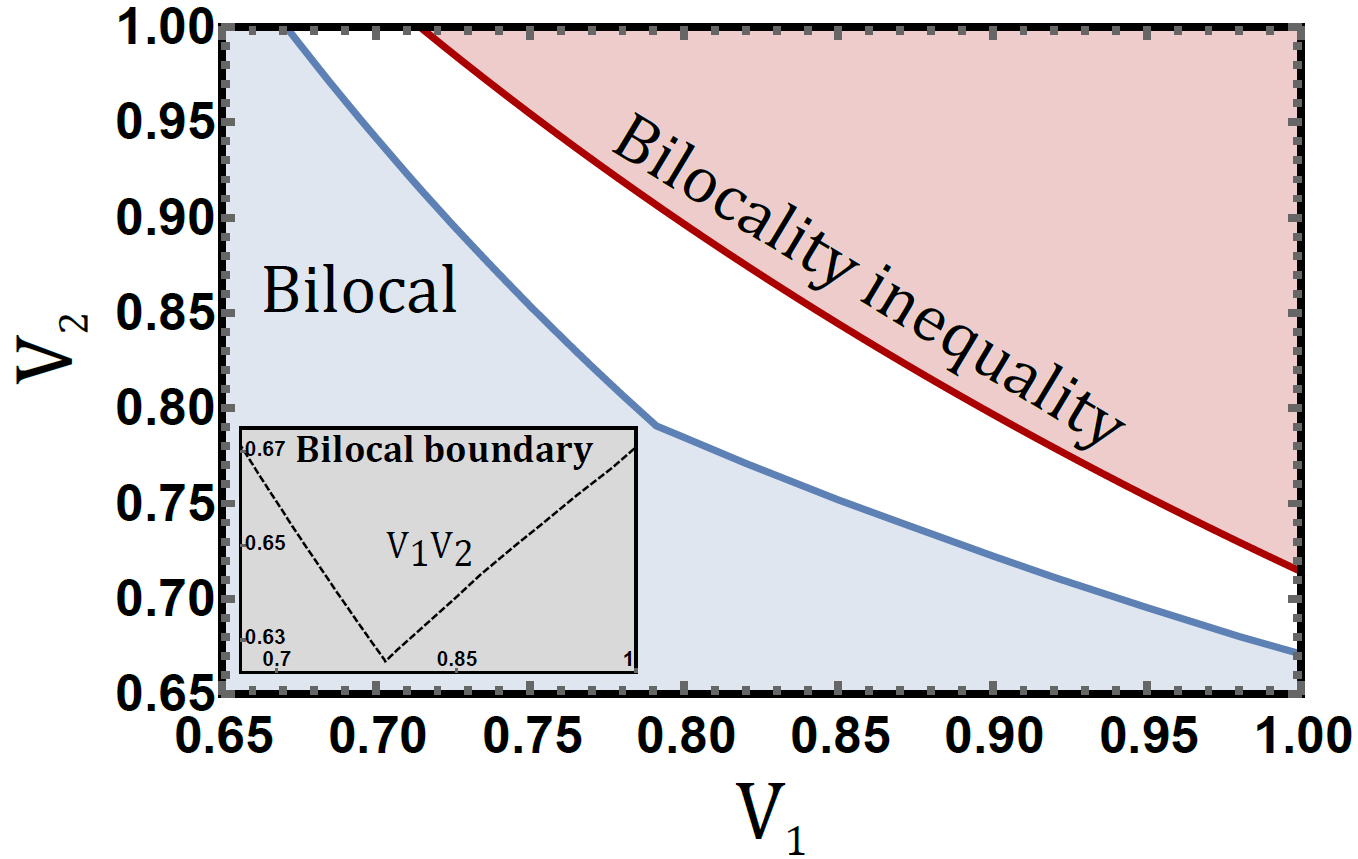}
	\caption{The blue region represents the set of bilocal quantum correlations $p_\text{Q}^{\theta=0}$ in the plane of visibilities $(V_1,V_2)$, with the dashed line in the inset figure showing the product of the visibilites on the boundary of this bilocal region. The red area is the part of the quantum region that can be detected as non-bilocal through the violation of our bilocal inequality~\eqref{bilocineq}.}\label{FigRegion}
\end{figure}

Secondly, we provide a better characterisation of the power of bilocal models by explicitly considering their ability to simulate the quantum correlations. To this end, we have used an efficient search method which exploits that the numerical difficulties associated with the bilocality assumption are significantly reduced if the bilocal model first undergoes a Fourier transformation~\cite{Branciard2}. For the case of $V\equiv V_1=V_2$ and $\theta = 0$ considered above, we look for the largest $V$ for which $p_\text{Q}^{\theta=0}$ admits a bilocal model, and find the critical visibility
\begin{equation}
V_\text{crit}\approx 79.1\,\%. \label{Vcrit_biloc_model}
\end{equation}
To further explore different values of $(V_1,V_2)$, we then also consider, for a given $V_1$, the largest $V_2$ for which a bilocal model exists. Figure~\ref{FigRegion} shows the region in the $(V_1,V_2)$-plane for which we find a bilocal simulation of $p_\text{Q}^\theta$ (still for $\theta = 0$ here; the analysis for $\theta > 0$ is presented in Appendix~\ref{AppEJMfamily}). It also displays the product $V_1 V_2$ associated to the boundary of the bilocal region (the critical pairs). Interestingly, the  product of the critical visibilities is not constant. This is in stark contrast with previously studied quantum correlations that arise from the BSM \cite{Branciard2} for which the product of visibilities determines the existence of a bilocal model. Notably, also the violations of many bilocal inequalities that are based on coordinated tests of standard Bell inequalities \cite{Branciard2, Tavakoli, Tavakoli2, Andreoli, Gisin} are determined by such products of visibilities. 

Thirdly, we employ an intuitive ansatz for analytically constructing bilocal models that mimic the symmetry of $p_\text{Q}^{\theta=0}$. Namely, we impose that the (unobserved) probability distribution $p_{\text{biloc}}(\alpha,b,\gamma)=q_\alpha^{(1)}q_\gamma^{(2)}p(b|\alpha,\gamma)$ of the bilocal model should have the same tetrahedral symmetry: for every permutation $\pi$ of the tetrahedron vertices $\{\vec m_b\}$ in Eq.~\eqref{tetra}, extended to the opposite vertices via $\pi(-\vec m_b) = -\pi(\vec m_b)$, and applied to the 3-vector variables $\alpha,b,\gamma$, one should have $p_{\text{biloc}}(\pi(\alpha),\pi(b),\pi(\gamma))=p_{\text{biloc}}(\alpha,b,\gamma)$. 
Under this symmetry ansatz, we are able to analytically construct efficient bilocal simulations of $p_\text{Q}^{\theta=0}$. Interestingly, along the entire boundary of the bilocal region, the obtained results match those presented in Figure~\ref{FigRegion} up to the fifth decimal digit. This shows that simple and highly symmetric bilocal models are very nearly optimal for simulating $p_\text{Q}^{\theta=0}$. These bilocal models and the critical visibilities are detailed in Appendix~\ref{AppSimulation}.

\textit{Bilocal Bell inequalities.---} We now draw inspiration from the structure of the nonbilocal quantum corelations obtained from the EJM to construct a bilocal inequality.  Hence, in contrast to several previous bilocal inequalities, the present one is neither based on, nor apparently resembles, a standard Bell inequality. Also, naturally, such an inequality applies to detecting the non-bilocality of general probability distributions, not only $p_\text{Q}^\theta$. To build the Bell expression, we introduce the following quantities
\begin{align} \label{Def_S_T}
S=\sum_{y=z}\,\langle B^yC_z\rangle - \sum_{x=y}\,\langle A_xB^y\rangle, \hspace{10mm} \nonumber \\[1mm]
T=\sum_{x\neq y\neq z\neq x}\langle A_xB^yC_z\rangle, \quad Z=\max \left(\mathcal{C}_\text{other}\right),
\end{align}
where $\mathcal{C}_\text{other}=\{|\langle A_x\rangle|, |\langle A_xB^y\rangle|, \ldots, |\langle A_xB^yC_z\rangle|\}$ is the set of the absolute values of all one-, two- and three-party correlators other than those appearing in the expressions of $S$ and $T$. This leads us to the following bilocal inequality:
\begin{equation}\label{bilocineq}
\mathcal{B}\equiv \frac{S}{3}-T \stackrel{\text{biloc}}{\leq} 3+5Z.
\end{equation}
Notice that the $Z$ quantity makes this general inequality nonlinear. The most interesting case is however when $Z=0$, as satisfied by the quantum correlation of Eq.~\eqref{correlators}. For this case, we have proved the bilocal bound under the  previously considered symmetry ansatz (which in fact enforces $Z=0$, see Appendix~\ref{AppTetraProof}). Then, we have also confirmed the bilocal bound using two different numerical methods applied to general bilocal models~\cite{NumericalFootNote}.  We find that the bilocal inequality above, for $Z=0$, is tight in the sense that it constitutes one of the facets of the projection of the ``$Z=0$ slice'' of the bilocal set of correlations onto the $(S,T)$-plane. Remarkably, this projection of the $Z=0$ slice is delimited by linear inequalities, as further described in Appendix~\ref{AppSlice}; this stands in contrast to previous bilocal inequalities which use nonlinear Bell expressions. Finally, for $Z>0$, we have again applied the same numerical search methods to justify the correction term $5Z$ in the bilocal bound of $\mathcal{B}$. Notably, more accurate corrections are also possible (see Appendix~\ref{AppCorrection}).

For our quantum correlation of Eq.~\eqref{correlators}, we straightforwardly obtain $(S,T,Z) = \left(3\frac{V_1+V_2}{2}\cos\theta,-3V_1V_2,0\right)$, and $\mathcal{B} = 3V_1V_2 + \frac{V_1+V_2}{2}\cos\theta$.
In the noiseless case ($V_1=V_2=1$), we thus get $\mathcal{B}=3 + \cos \theta$, which gives a violation of our bilocal inequality~\eqref{bilocineq} for our whole family of generalised EJM (i.e., the whole range of $\theta$), except for the special case of a BSM ($\theta=\frac{\pi}{2}$, for which our quantum correlation turns out to be bilocal: see Appendix~\ref{AppEJMfamily}).
In contrast, when white noise is present and both sources are equally noisy ($V\equiv V_1=V_2$), we get a violation of our inequality whenever $3V^2 + V \cos\theta > 3$. For $\theta = 0$, the critical visibility per singlet required for a violation is
\begin{equation} \label{Vcrit_ineq}
V_\text{crit}=\frac{\sqrt{37}-1}{6}\approx 84.7\,\%.
\end{equation}
This shows that the quantum violation is robust to white noise on the singlet states, but not optimally robust as no violation is found here for $V\in[0.791,0.847]$. More generally, the bilocal inequality enables the detection of quantum correlations in a sizable segment of the $(V_1,V_2)$-plane (see Figure~\ref{FigRegion}).

Finally, we note that several different bilocal inequalities can be constructed based on the correlations from the EJM. As another example, in Appendix~\ref{AppSecondIneq} we consider the following Bell expression
\begin{align}\nonumber \label{eq:SecondIneq}
\mathcal{B}' \equiv & \sum_{x, b}\sqrt{p(b)\left(1-b^x E^\text{A}_b(x)\right)}+\sum_{z, b}\sqrt{p(b)\left(1+b^z E^\text{C}_b(z)\right)} \\
& \quad + \sum_{x\neq z, b}\sqrt{p(b)\left(1-b^x b^z E^\text{AC}_b(x,z)\right)},
\end{align}
where $E^\text{A}_b(x)$, $E^\text{C}_b(z)$ and $E^\text{AC}_b(x,z)$ denote one- and two-party expectation values for Alice and Charlie, conditioned on Bob's output $b = (b^1,b^2,b^3)$ (see Appendix~\ref{AppSecondIneq}). Numerical methods similar to the previous ones are employed to evidence that $\mathcal{B}'\leq 12\sqrt{3} + 2\sqrt{15}$ holds for bilocal models. In Appendix~\ref{AppSecondIneq} we prove that there are quantum distributions whose non-bilocality is detected with this bilocal inequality but not with the inequality~\eqref{bilocineq}. Furthermore, we also prove that if Bob has uniformly distributed outcomes ($p(b)=\frac{1}{4}$), then  $\mathcal{B}' \lesssim 30.70$ is respected by all quantum models with independent sources and hence it constitutes a quantum Bell inequality for the network~\cite{Pozas2}.

\textit{Implementation of the Elegant Joint Measurement.---} It is both interesting and practically relevant to address the question of how one may implement experimentally the EJM and its generalisation. In general, the implementation of joint (two-qubit) measurements requires the interaction of different signals. Optical implementations are of particular interest  since they are common and convenient for Bell-type experiments. However, many such measurements, including the BSM, cannot be implemented with the basic tools applied in linear optics schemes (phase-shifters and beam splitters) when no auxiliary photons are present~\cite{Lutkenhaus}. It turns out that our family of generalised EJM as defined by Eq.~~\eqref{Phi_b} can also not be implemented with two-photon linear optics, as can be shown by evaluating the criterion provided in Ref.~\cite{Loock}. More sophisticated tools are therefore required.

\begin{figure}[t!]
	\centering
	\includegraphics[width=0.9\columnwidth]{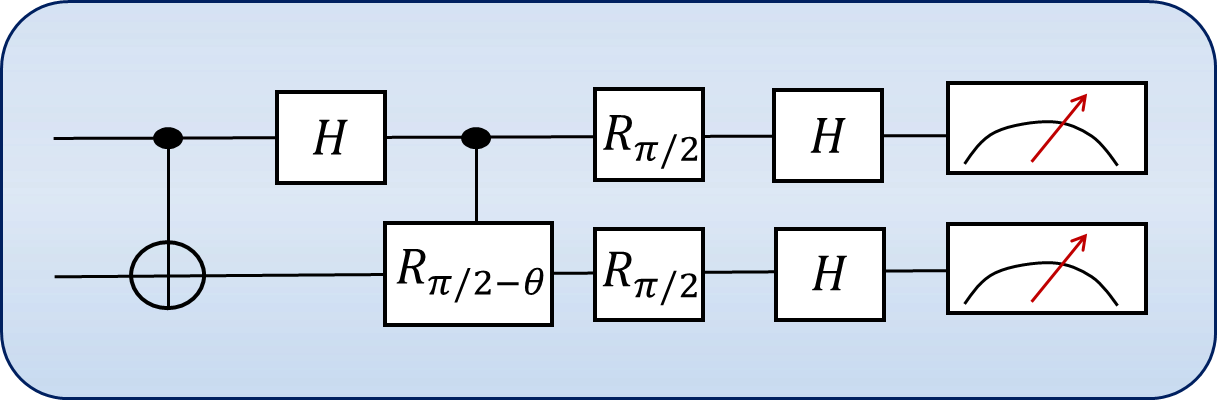}
	\caption{Quantum circuit for implementing our family of generalised Elegant Joint Measurements parameterised by $\theta$. A controlled-NOT gate is followed by a Hadamard rotation ($H=(\sigma_1+\sigma_3)/\sqrt{2}$) on the control qubit, a controlled phase shift gate $R_{\pi/2-\theta}$, and a separate rotation of each qubit composed of $R_{\pi/2}$ and $H$. Finally, a measurement is performed in the basis $\{\ket{00},\ket{01},\ket{10},\ket{11}\}$.}\label{FigCircuit}
\end{figure}

Our measurement family can in fact be implemented by the two-qubit circuit presented in Figure~\ref{FigCircuit}. This circuit maps the four measurement basis states $\{\ket{\Phi_b^\theta}\}_b$ onto the computational basis product states $\{\ket{00},\ket{01},\ket{10},\ket{11}\}$ (up to global phases). The proposed  implementation involves (in addition to single-qubit gates) two different controlled unitary operations, namely a standard controlled-NOT gate and a controlled implementation of the phase shift gate 
\begin{equation}
R_\phi=\begin{pmatrix}
1 & 0\\
0 & e^{i\phi}
\end{pmatrix}.
\end{equation}
We remark that this controlled phase gate itself can be implemented using two controlled-NOT gates and unitaries acting on the target qubit as described in Ref.~\cite{Barenco}.
Finally, notice that when $\theta=\frac{\pi}{2}$, we have $R_{\pi/2-\theta}=\openone$ and thus the circuit only involves a single two-qubit gate, just like the standard scheme for a BSM \cite{Nielsen}.

\textit{Discussion and open questions.---}	We have investigated quantum violations of bilocality based on the Elegant Joint Measurement and a new generalisation thereof. In contrast to several previous works in which quantum correlations were generated through a Bell State Measurement, our setup does not effectively reduce to separate implementations of the standard CHSH scenario. We nevertheless constructed new bilocal inequalities, and exhibited violations that we could not directly trace back to violations of a standard Bell inequality. Finally, we paved the way towards a bilocality experiment based on the EJM by constructing a quantum circuit for its implementation.

Several intriguing questions are left open. 1) What is the largest possible quantum violation of the bilocal inequalities? 2) Can the inequalities be proven in full generality? We note that the semidefinite relaxation methods of~\cite{Pozas} can be exploited to place a bilocal bound on $\mathcal{B}$, albeit perhaps not tight. 3) How can one formalise the intuitive idea that some bilocal inequalities may or may not trace back to standard Bell inequalities? 4) Can our EJM family be further generalised for two higher-dimensional systems or for more than two qubits such that it preserves its elegant properties? 5) Are there any other correlations obtained using our EJM family that would be of particular interest to study (in the bilocality scenario or beyond), and more generally, could the introduced family of measurements have other interesting applications in quantum information science?

\begin{acknowledgements}
We thank Alejandro Pozas-Kerstjens for sharing with us both his codes  and results for semidefinite programs based on Ref.~\cite{Pozas}, and Norbert L\"utkenhaus for directing us into Ref.~\cite{Loock}. This work was supported by the Swiss National Science Foundation via the NCCR-SwissMap.
\end{acknowledgements}

\appendix

\newpage
\section{Bilocal simulation for intermediate measurements}\label{AppEJMfamily}	
Here, we explore the possibility of a bilocal simulation of quantum correlations based on the measurement family intermediate between the EJM and the BSM. Firstly, we consider the case in which Alice and Charlie perform the measurements $(\sigma_1,\sigma_2,\sigma_3)$ and Bob performs the intermediate measurement corresponding to a fixed $\theta$. The resulting correlators are given in Eq.~\eqref{correlators} of the main text. 

\begin{figure}
	\centering
	\includegraphics[width=\columnwidth]{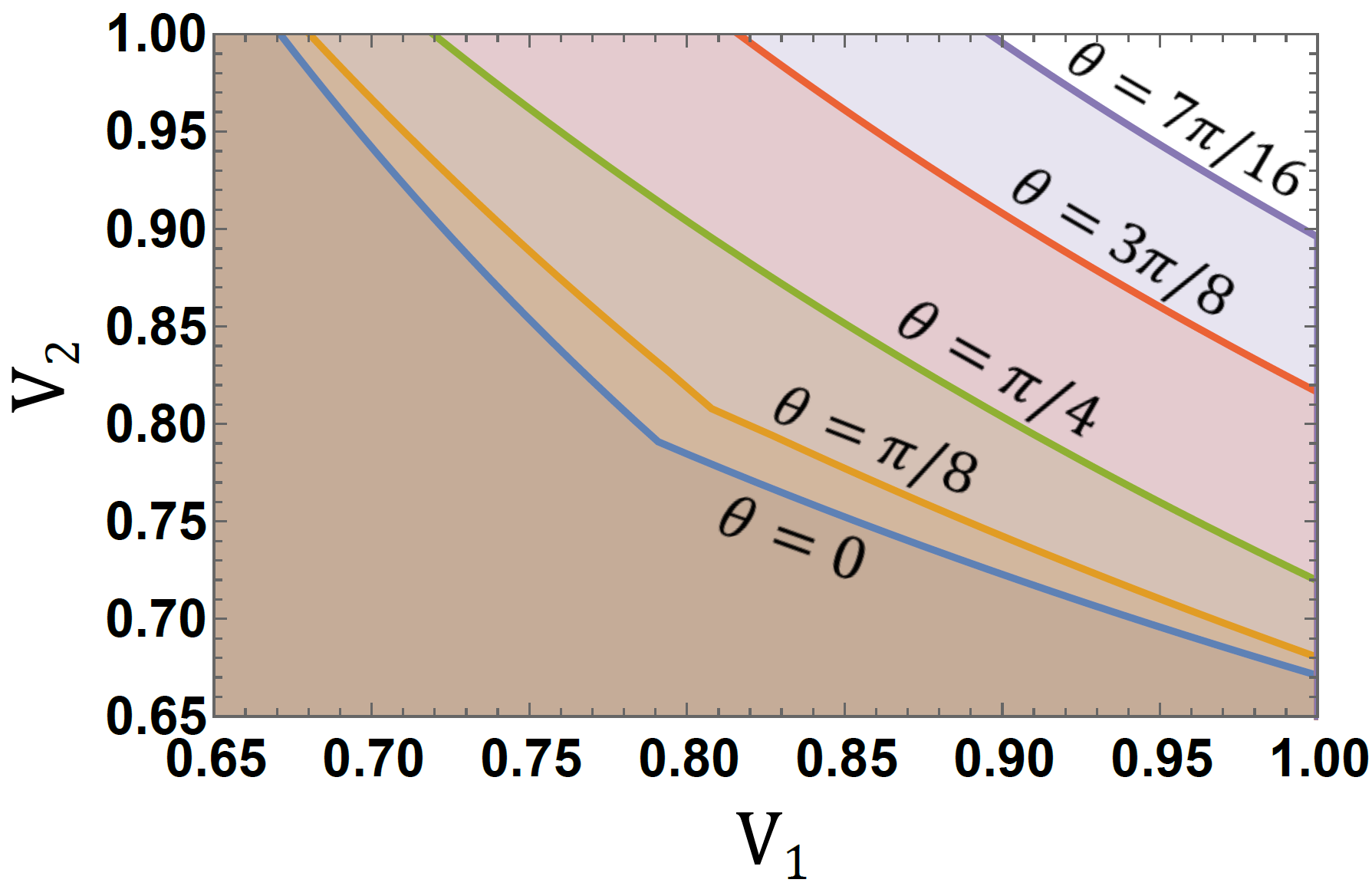}
	\caption{Bilocal regions of the quantum correlations $p_\text{Q}^\theta$ in the $(V_1,V_2)$-plane, for different values of $\theta$.}\label{FigIntermediateSimulation}
\end{figure}

In order to search for the region in the $(V_1,V_2)$-plane for which a bilocal simulation is possible, we have used the numerical method mentioned in the main text (where we presented the analysis for $\theta=0$). Specifically, for a given $\theta$, we search for a brute-force solution to $p_\text{Q}^\theta=p_\text{biloc}$  where we first apply a Fourier transform to the problem. This transforms probabilities into correlators. Some of these correlators are fixed immediately by the constraint $p_\text{Q}^\theta=p_\text{biloc}$. Those that are not fixed correspond to non-observable correlators (say e.g.~$\langle A_1A_2C_1\rangle$) and represent the internal degrees of freedom of the bilocal model, which we optimise over (under the constraint that they define valid probabilities). The main benefit of this method is that source-independence, appearing on the level of the free correlators, translates into simple conditions that are either linear or quadratic. This makes the numerical search more efficient and accurate; see Ref.~\cite{Branciard2} for a more detailed description. In Figure~\ref{FigIntermediateSimulation}, we display the boundary of the bilocal region found through this method for several different values of $\theta$. We find that as we depart further from the EJM, i.e~as we increase $\theta$, the region that admits a bilocal simulation grows larger. For $\theta = \frac{\pi}{2}$, the quantum correlation $p_\text{Q}^\theta$ is found to be bilocal\footnote{An explicit bilocal model for $p_\text{Q}^{\theta = \frac{\pi}{2}}$ is obtained by letting $\alpha$ be any of the 4 vectors $-\vec m_b$ and $\gamma$ be any of the 4 vectors $\vec m_b$ of Eq.~\eqref{tetra}, with equal probabilities, and by defining $p(b|\alpha,\gamma) = \frac{1+3 V_1 V_2}{4}$ if $-\alpha=b =\gamma$ or $\det(-\alpha,b,\gamma) > 0$, and $p(b|\alpha,\gamma) = \frac{1-V_1 V_2}{4}$ otherwise.}  \label{footnote_biloc_model_theta_pi_2} for all visibilities $V_1,V_2$.

\begin{figure}
	\centering
	\includegraphics[width=\columnwidth]{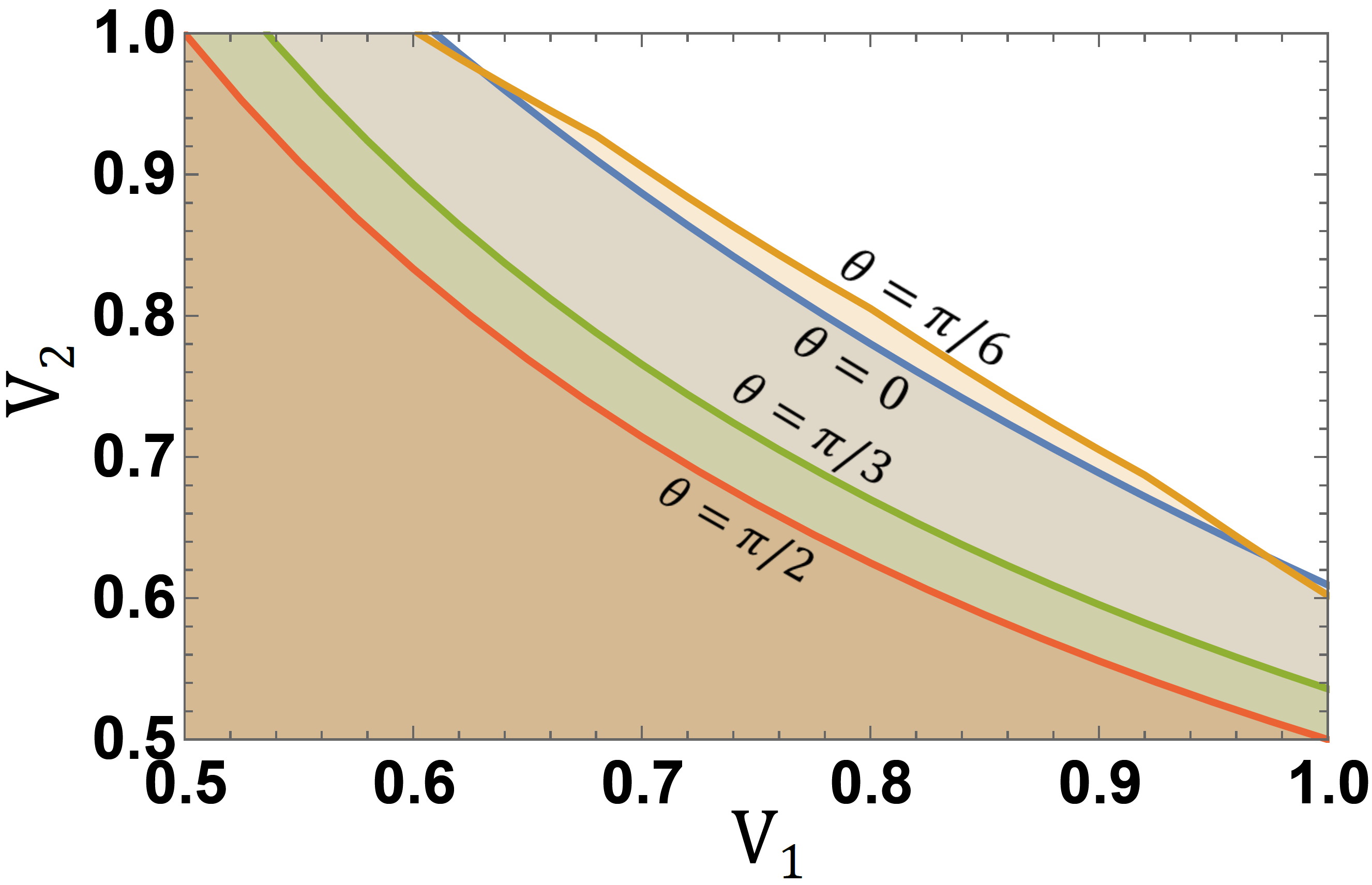}
	\caption{Bilocal regions of the quantum correlations when Alice and Charlie perform the measurements $\big(\frac{\sigma_3+\sigma_1}{\sqrt{2}},\sigma_2,\frac{\sigma_3-\sigma_1}{\sqrt{2}}\big)$ for different values of $\theta$.}\label{FigIntermediateSimulation2}
\end{figure}

It is interesting to note that while the non-bilocal region appears to be vanishing in Figure~\ref{FigIntermediateSimulation} as Bob's measurement approaches the BSM, the standard bilocal inequality~\cite{Branciard2}, which is based on the BSM, admits a robust quantum violation. This suggests that as $\theta$ grows larger, and the measurement becomes less similar to the EJM and more similar to the BSM, Alice and Charlie would benefit from changing the orientation of their local measurements. We illustrate this by letting Alice and Charlie measure in the bases $\big(\frac{\sigma_3+\sigma_1}{\sqrt{2}},\sigma_2,\frac{\sigma_3-\sigma_1}{\sqrt{2}}\big)$. For several values of $\theta$, we plot the region in the $(V_1,V_2)$-plane for which we find a bilocal simulation of the quantum correlations thus obtained (still considering measurements on noisy singlet states): see Figure~\ref{FigIntermediateSimulation2}. In Figure~\ref{FigIntermediateSimulation2} we see that the trend observed in Figure~\ref{FigIntermediateSimulation} is reversed; for larger values of $\theta$, the bilocal region is shrinking. In particular, for the BSM ($\theta=\frac{\pi}{2}$), the boundary of the bilocal region is characterised by $V_1V_2=\frac{1}{2}$ which is the same as that obtained in the standard bilocal inequality~\cite{Branciard2}. However, the bilocal region in Figure~\ref{FigIntermediateSimulation2} is not monotonic in $\theta$: the bilocal region for $\theta=\frac{\pi}{6}$ is typically larger than that of the EJM ($\theta=0$). Typically, for small values of $\theta$, the re-oriented local measurements of Alice and Charlie do not constitute an improvement over the previous $(\sigma_1,\sigma_2,\sigma_3)$ measurements.

All this illustrates the fact that the choice of Alice and Charlie's measurements have a nontrivial implication on the existence or non-existence of a bilocal model for the quantum correlations under investigation. Although the choice of measurements $(\sigma_1,\sigma_2,\sigma_3)$ for Alice and Charlie that we considered in the main text looks appropriate when Bob performs the EJM, it is seen to be nonoptimal when Bob uses the generalised EJM family, for general values of $\theta > 0$. Finding the optimal measurements to unveil quantum nonbilocality in a given scenario is certainly not a trivial task.

\section{Bilocal models with tetrahedral symmetry}\label{AppSimulation}

We detail here a simple and analytical family of bilocal models exhibiting the tetrahedral symmetry outlined in the main text. These models provide bilocal simulations of the quantum correlation $p_\text{Q}^{\theta=0}$ for visibilities $(V_1,V_2)$ very close to the critical ones, above which $p_\text{Q}^{\theta=0}$ becomes nonbilocal. Along the boundary of the bilocal set in the $(V_1,V_2)$ plane (shown in Figure~\ref{FigRegion} of the main text), the difference between the critical pairs obtained by numerical search without our symmetry assumption and those obtained for the model below is of the order of $10^{-5}$ only; e.g., for the symmetric noise case ($V_1=V_2$), numerical optimisation over all bilocal models gave us a critical visibility of $V_\text{crit}\approx 0.790896$, while the symmetric model below gives $V_\text{crit}\approx 0.790871$. 
	
For convenience in the presentation below, let us denote by ${\cal T}_+ = \{\vec m_b\}_{b= 1,\ldots,4}$ the set of tetrahedron vertices $\vec m_b$ of Eq.~\eqref{tetra} and by ${\cal T}_- = \{-\vec m_b\}_{b= 1,\ldots,4}$ the set of opposite vectors. For any $\alpha = (\alpha_1,\alpha_2,\alpha_3) \in {\cal T}_\pm$ (for any $\gamma = (\gamma_1,\gamma_2,\gamma_3) \in {\cal T}_\pm$, respectively), let us define $\tilde\alpha = \pm \alpha \in {\cal T}_+$ ($\tilde\gamma = \pm \gamma \in {\cal T}_+$) to be the vector in ${\cal T}_+$ along the same direction as $\alpha$ ($\gamma$) and possibly with the opposite sign, if $\alpha\in {\cal T}_-$ (if $\gamma \in {\cal T}_-$).
	
In general, $p_\text{Q}^\theta=p_\text{biloc}$ leads to a large system of equations. However, the symmetry ansatz greatly simplifies the problem. 
Note first that the requirement that $p_{\text{biloc}}(\pi(\alpha),\pi(b),\pi(\gamma))=p_{\text{biloc}}(\alpha,b,\gamma)$ for all permutations $\pi$ of the tetrahedron (as defined in the main text) imposes that the probabilities $q_\alpha^{(1)}$ ($q_\gamma^{(2)}$, resp.) are the same for all four values of $\alpha$ ($\gamma$) in ${\cal T}_+$, and the same for all four values of $\alpha$ ($\gamma$) in ${\cal T}_-$. Defining $q_\pm^{(1)} = \sum_{\alpha\in{\cal T}_\pm} q_{\alpha}^{(1)} \in [0,1]$ and $q_\pm^{(2)} = \sum_{\gamma\in{\cal T}_\pm} q_{\gamma}^{(2)} \in [0,1]$, the weights $q^{(1)}_\alpha$ ($q^{(2)}_\gamma$) are then all either equal to $\frac14 q^{(1)}_+$ ($\frac14 q^{(2)}_+$) or to $\frac14 q^{(1)}_-$ ($\frac14 q^{(2)}_-$), depending on whether $\alpha \ (\gamma)$ is in ${\cal T}_+$ or ${\cal T}_-$.

In turn, it also follows that Bob's response functions conditioned on the local variables $\alpha, \gamma$ have	 the symmetry $p(\pi(b)|\pi(\alpha),\pi(\gamma))=p(b|\alpha,\gamma)$ for all permutations $\pi$ of the tetrahedron. With this symmetry (and noting that $b$, just like $\tilde\alpha$ and $\tilde\gamma$, is always in ${\cal T}_+$), Bob's response functions can be defined by only specifying for instance, for each of the four cases where $\alpha \in {\cal T}_\pm$ and $\gamma \in {\cal T}_\pm$: \emph{(i)} the probabilities that $b=\tilde\alpha=\tilde\gamma$ when $\tilde\alpha=\tilde\gamma$, which we denote by $q^{\tau_\alpha,\tau_\gamma}_{b=\tilde\alpha=\tilde\gamma|\tilde\alpha=\tilde\gamma}$ (with the superscripts $\tau_\alpha,\tau_\gamma=\pm$ referring to $\alpha \in {\cal T}_{\tau_\alpha}$ and $\gamma \in {\cal T}_{\tau_\gamma}$, and such that the probabilities that $b$ takes any of the three values other than $\tilde\alpha=\tilde\gamma$ is, by symmetry, $(1-q^{\tau_\alpha,\tau_\gamma}_{b=\tilde\alpha=\tilde\gamma|\tilde\alpha=\tilde\gamma})/3$);  \emph{(ii)} the probabilities that $b=\tilde\alpha$ and \emph{(iii)} the probabilities that $b=\tilde\gamma$ when $\tilde\alpha\neq\tilde\gamma$, which we denote by $q^{\tau_\alpha,\tau_\gamma}_{b=\tilde\alpha|\tilde\alpha\neq\tilde\gamma}$ and $q^{\tau_\alpha,\tau_\gamma}_{b=\tilde\gamma|\tilde\alpha\neq\tilde\gamma}$, resp. (such that the probabilities that $b$ takes any of the two values other than $\tilde\alpha$ and $\tilde\gamma$, when these are different, is $(1-q^{\tau_\alpha,\tau_\gamma}_{b=\tilde\alpha|\tilde\alpha\neq\tilde\gamma}-q^{\tau_\alpha,\tau_\gamma}_{b=\tilde\gamma|\tilde\alpha\neq\tilde\gamma})/2$). All in all (and noting that $q_-^{(i)} = 1 - q_+^{(i)}$ for $i=1,2$), any bilocal model with the tetrahedral symmetry considered here can thus be defined by just the 14 parameters $q_+^{(1)}, q_+^{(2)}, q^{\tau_\alpha,\tau_\gamma}_{b=\tilde\alpha=\tilde\gamma|\tilde\alpha=\tilde\gamma}, q^{\tau_\alpha,\tau_\gamma}_{b=\tilde\alpha|\tilde\alpha\neq\tilde\gamma}, q^{\tau_\alpha,\tau_\gamma}_{b=\tilde\gamma|\tilde\alpha\neq\tilde\gamma}$ (for each of the four combinations of $\tau_\alpha,\tau_\gamma$).%
\footnote{Note that $p_\text{Q}^{\theta}$ does not have the (``full'') tetrahedral symmetry considered here when $\theta > 0$, as the correlators $\langle A_xB^yC_z\rangle$ in Eq.~\eqref{correlators} are different for even and odd permutations of $\{x,y,z\}$. One may also define bilocal models with a ``relaxed'' tetrahedral symmetry matching the symmetry of $p_\text{Q}^{\theta>0}$, by allowing for different probabilities for the two values of $b$ other than $\tilde\alpha$ and $\tilde\gamma$ when these are different (just depending on the sign of $\det(\tilde\alpha,b,\tilde\gamma)$ to preserve some symmetry). This would add four parameters to the model (one for each combination of $\tau_\alpha,\tau_\gamma$). As an example, the explicit bilocal model given in Footnote~\ref{footnote_biloc_model_theta_pi_2} for $p_\text{Q}^{\theta=\frac{\pi}{2}}$ has this relaxed tetrahedral symmetry. In the remaining part of these appendices however, by tetrahedral symmetry we will refer to the ``full'' tetrahedral symmetry.} \label{footnote_relaxed_tetra_sym}

To find the critical visibilities $(V_1,V_2)$ for which such symmetric models can reproduce the quantum correlation $p_\text{Q}^{\theta=0}$, we let $V_1$ take different fixed values, and optimise	 over the 14 weights above, together with $V_2$, so as to find the largest possible $V_2$ allowing for $p_\text{Q}^{\theta=0}$ to be reproduced. Numerically we found, for large enough $V_1$ (namely, $V_1 \gtrsim 0.791$), that the optimal strategies were to take
\begin{align}
& q_+^{(1)} \approx q_+^{(2)}, \nonumber \\
& q^{+,+}_{b=\tilde\alpha=\tilde\gamma|\tilde\alpha=\tilde\gamma} \approx 0, \quad q^{+,+}_{b=\tilde\alpha|\tilde\alpha\neq\tilde\gamma} \approx 0, \quad q^{+,+}_{b=\tilde\gamma|\tilde\alpha\neq\tilde\gamma} \approx 1, \nonumber \\
& q^{-,+}_{b=\tilde\alpha=\tilde\gamma|\tilde\alpha=\tilde\gamma} \approx 1, \quad q^{-,+}_{b=\tilde\alpha|\tilde\alpha\neq\tilde\gamma} \approx 0, \quad q^{-,+}_{b=\tilde\gamma|\tilde\alpha\neq\tilde\gamma} = q_0, \nonumber \\
& q^{+,-}_{b=\tilde\alpha=\tilde\gamma|\tilde\alpha=\tilde\gamma} \approx 1, \quad q^{+,-}_{b=\tilde\alpha|\tilde\alpha\neq\tilde\gamma} \approx 0, \quad q^{+,-}_{b=\tilde\gamma|\tilde\alpha\neq\tilde\gamma} \approx 0, \nonumber \\
& q^{-,-}_{b=\tilde\alpha=\tilde\gamma|\tilde\alpha=\tilde\gamma} \approx 0, \quad q^{-,-}_{b=\tilde\alpha|\tilde\alpha\neq\tilde\gamma} \approx 1, \quad q^{-,-}_{b=\tilde\gamma|\tilde\alpha\neq\tilde\gamma} \approx 0, \label{eq:symmetric_optim_weights}
\end{align}
for some value $q_0 \in [0,1]$ (that depends on $V_1$). E.g., when $\alpha,\gamma\in{\cal T}_+$ (in which case $\tilde\alpha=\alpha, \tilde\gamma=\gamma$), then the model should return any of the three values $b \neq \alpha,\gamma$ (with equal probabilities) if $\alpha=\gamma$, or should return $b = \gamma$ if $\alpha\neq\gamma$; when $\alpha \in {\cal T}_-, \gamma\in{\cal T}_+$ (in which case $\tilde\alpha=-\alpha, \tilde\gamma=\gamma$), then the model should return $b = -\alpha = \gamma$ if $-\alpha=\gamma$, or should return $b = \gamma$ with probability $q_0$ or any of the two values $b \neq -\alpha,\gamma$ with equal probabilities $(1-q_0)/2$ if $-\alpha\neq\gamma$; etc.

By imposing that the 14 parameters of the model satisfy Eq.~\eqref{eq:symmetric_optim_weights} with strict equalities, it becomes possible to construct the model analytically. To reproduce the correlation $p_\text{Q}^{\theta=0}$, for a given value of $V_1$, the remaining free parameters need to take the values
\begin{align}
& V_2 = \frac{58+9V_1-12\sqrt{2V_1 - 8/9}}{27(1+2V_1)} , \nonumber \\[1mm]
& q_+^{(1)} = q_+^{(2)} = \frac23 - \frac{\sqrt{2V_1 - 8/9}}{2}, \nonumber \\[1mm]
& q_0 = \frac{6\sqrt{2V_1 - 8/9}+9 V_2-9 V_1-2}{3\sqrt{2V_1 - 8/9}+8-9 V_1}, \label{remaining_params_sym_model}
\end{align}
which completes the full specification of our family of bilocal models for $p_\text{Q}^{\theta=0}$, and for some very close-to-optimal visibilities $(V_1,V_2)$.

Note that our models here work for visibilities $V_1 \ge V_2$; for $V_2 \ge V_1$ similar models can be found, with the roles of $V_1$ and $V_2$ exchanged in the construction above.
For $V_1 = V_2$, the critical visibility $V_\text{crit}\approx 0.791$ is obtained as the unique solution to the first line of Eq.~\eqref{remaining_params_sym_model}, after imposing $V_1 = V_2 = V_\text{crit}$. Note also that $V_1 \ge V_2$ ensures in particular that $V_1 \ge V_\text{crit} > 4/9$, so that the square roots in Eq.~\eqref{remaining_params_sym_model} take real values.

\section{Proof of the first bilocal inequality under tetrahedral symmetry}\label{AppTetraProof}	

Here we prove the bilocal inequality~\eqref{bilocineq} for models that satisfy our tetrahedral symetry ansatz.

For such models, as defined in Appendix~\ref{AppSimulation} in terms of the 14 parameters $q_+^{(1)}, q_+^{(2)}, q^{\tau_\alpha,\tau_\gamma}_{b=\tilde\alpha=\tilde\gamma|\tilde\alpha=\tilde\gamma}, q^{\tau_\alpha,\tau_\gamma}_{b=\tilde\alpha|\tilde\alpha\neq\tilde\gamma}, q^{\tau_\alpha,\tau_\gamma}_{b=\tilde\gamma|\tilde\alpha\neq\tilde\gamma}$, the one-, two- and three-party correlators are found to be
\begin{align}\label{correlators_sym_model}\nonumber
& \langle A_x\rangle = \langle B^y\rangle = \langle C_z\rangle = \langle A_xC_z\rangle = 0, \nonumber \\[1mm]
& \langle A_xB^y\rangle= \delta_{x,y} \!\!\!\! \sum_{\tau_\alpha,\tau_\gamma=\pm1} \!\!\!\! q_{\tau_\alpha}^{(1)}q_{\tau_\gamma}^{(2)} \, \tau_\alpha \Big( q^{\tau_\alpha,\tau_\gamma}_{b=\tilde\alpha|\tilde\alpha\neq\tilde\gamma} - {\textstyle \frac{1-q^{\tau_\alpha,\tau_\gamma}_{b=\tilde\alpha=\tilde\gamma|\tilde\alpha=\tilde\gamma}}{3}}\Big), \nonumber \\[1mm]
& \langle B^yC_z\rangle= \delta_{y,z} \!\!\!\! \sum_{\tau_\alpha,\tau_\gamma=\pm1} \!\!\!\! q_{\tau_\alpha}^{(1)}q_{\tau_\gamma}^{(2)} \, \tau_\gamma \Big( q^{\tau_\alpha,\tau_\gamma}_{b=\tilde\gamma|\tilde\alpha\neq\tilde\gamma} - {\textstyle \frac{1-q^{\tau_\alpha,\tau_\gamma}_{b=\tilde\alpha=\tilde\gamma|\tilde\alpha=\tilde\gamma}}{3}}\Big), \nonumber \\[1mm]
& \langle A_xB^yC_z\rangle= \delta_{x\neq y\neq z} \!\!\!\! \sum_{\tau_\alpha,\tau_\gamma=\pm1} \!\!\!\! q_{\tau_\alpha}^{(1)}q_{\tau_\gamma}^{(2)} \, \tau_\alpha\tau_\gamma \Big( {\textstyle \frac12 - \frac{1-q^{\tau_\alpha,\tau_\gamma}_{b=\tilde\alpha=\tilde\gamma|\tilde\alpha=\tilde\gamma}}{3}} \nonumber \\[-3mm]
& \hspace{55mm} - {\textstyle \frac{q^{\tau_\alpha,\tau_\gamma}_{b=\tilde\alpha|\tilde\alpha\neq\tilde\gamma} + q^{\tau_\alpha,\tau_\gamma}_{b=\tilde\gamma|\tilde\alpha\neq\tilde\gamma}}{2}} \Big)
\end{align}
(with $\delta_{x\neq y\neq z} = 1$ if $x,y,z$ are all distinct, $\delta_{x\neq y\neq z} = 0$ otherwise). From these we get%
\footnote{Note that bilocal models with the ``relaxed'' tetrahedral symmetry as described in Footnote~\ref{footnote_relaxed_tetra_sym} give the same values of $S$ and $T$, so that the proof here also applies to such models.}
\begin{align}
& \hspace{-5mm} \frac{S}{3}-T \nonumber \\
= \ \ & q_+^{(1)}q_+^{(2)} \Big[ 2 q^{+,+}_{b=\tilde\alpha|\tilde\alpha\neq\tilde\gamma} + 4 q^{+,+}_{b=\tilde\gamma|\tilde\alpha\neq\tilde\gamma} - 1 - 2q^{+,+}_{b=\tilde\alpha=\tilde\gamma|\tilde\alpha=\tilde\gamma} \Big] \nonumber \\
+ \, & q_+^{(1)}q_-^{(2)} \Big[ 3-4{\textstyle \frac{1-q^{+,-}_{b=\tilde\alpha=\tilde\gamma|\tilde\alpha=\tilde\gamma}}{3}} - 4q^{+,-}_{b=\tilde\alpha|\tilde\alpha\neq\tilde\gamma} - 4q^{+,-}_{b=\tilde\gamma|\tilde\alpha\neq\tilde\gamma} \Big] \nonumber \\
+ \, & q_-^{(1)}q_+^{(2)} \Big[ 3-8{\textstyle \frac{1-q^{-,+}_{b=\tilde\alpha=\tilde\gamma|\tilde\alpha=\tilde\gamma}}{3}} - 2q^{-,+}_{b=\tilde\alpha|\tilde\alpha\neq\tilde\gamma} - 2q^{-,+}_{b=\tilde\gamma|\tilde\alpha\neq\tilde\gamma} \Big] \nonumber \\
+ \, & q_-^{(1)}q_-^{(2)} \Big[ 4 q^{-,-}_{b=\tilde\alpha|\tilde\alpha\neq\tilde\gamma} +2 q^{-,-}_{b=\tilde\gamma|\tilde\alpha\neq\tilde\gamma} - 1 - 2q^{-,-}_{b=\tilde\alpha=\tilde\gamma|\tilde\alpha=\tilde\gamma} \Big] \label{B_sym_model}
\end{align}
and $Z=0$.

Recalling that all parameters $q_{(\cdots)}^{\tau_\alpha,\tau_\gamma}$ of the symmetric model are between $0$ and $1$, and that they further satisfy $q^{\tau_\alpha,\tau_\gamma}_{b=\tilde\alpha|\tilde\alpha\neq\tilde\gamma} + q^{\tau_\alpha,\tau_\gamma}_{b=\tilde\gamma|\tilde\alpha\neq\tilde\gamma} \le 1$, one can easily see that each term in square brackets above is upper-bounded by $3$. As $S/3-T$ is obtained as a convex combination of these four terms (with the weights $q_\pm^{(1)}q_\pm^{(2)}$), then it is also upper-bounded by 3---which indeed proves our inequality~\eqref{bilocineq} for bilocal models satisfying the tetrahedral symmetry assumption (for which $Z=0$).

We believe it should be possible to prove that any general bilocal model for correlations satisfying $Z=0$ can be ``symmetrised'' into a bilocal model with the tetrahedral symmetry considered here, that would have the same values of $S$ and $T$. This would give a general proof of our bilocal inequality~\eqref{bilocineq}, for the $Z=0$ case. However, the details here remain to be worked out properly, so that we rely for now on (trustworthy) numerical optimisations.

\section{``$Z=0$ slice'' of the bilocal set in the $(S,T)$-plane}\label{AppSlice}

It clearly appears that a case of particular interest in our study is when $Z=0$---as satisfied in particular by the quantum correlation $p_\text{Q}^{\theta}$ we investigate, and by any bilocal model with the tetrahedral symmetry considered previously. The choice to define and look at the quantities $S$ and $T$, as defined in Eq.~\eqref{Def_S_T}, is then rather naturally dictated by the specific forms of the correlators, Eq.~\eqref{correlators} for $p_\text{Q}^{\theta}$.%
\footnote{One may also naturally refine the analysis by defining and considering $S^{AB} = \sum_{x=y}\,\langle A_xB^y\rangle$, $S^{BC} =\sum_{y=z}\,\langle B^yC_z\rangle$, $R^+ = \sum_{xyz\in\{123,231,312\}}\langle A_xB^yC_z\rangle$ and $R^- = \sum_{xyz\in\{132,213,321\}}\langle A_xB^yC_z\rangle$ (such that $S = S^{BC} - S^{AB}$ and $R = R^+ + R^-$).}

\begin{figure}
	\centering
	\includegraphics[width=.8\columnwidth]{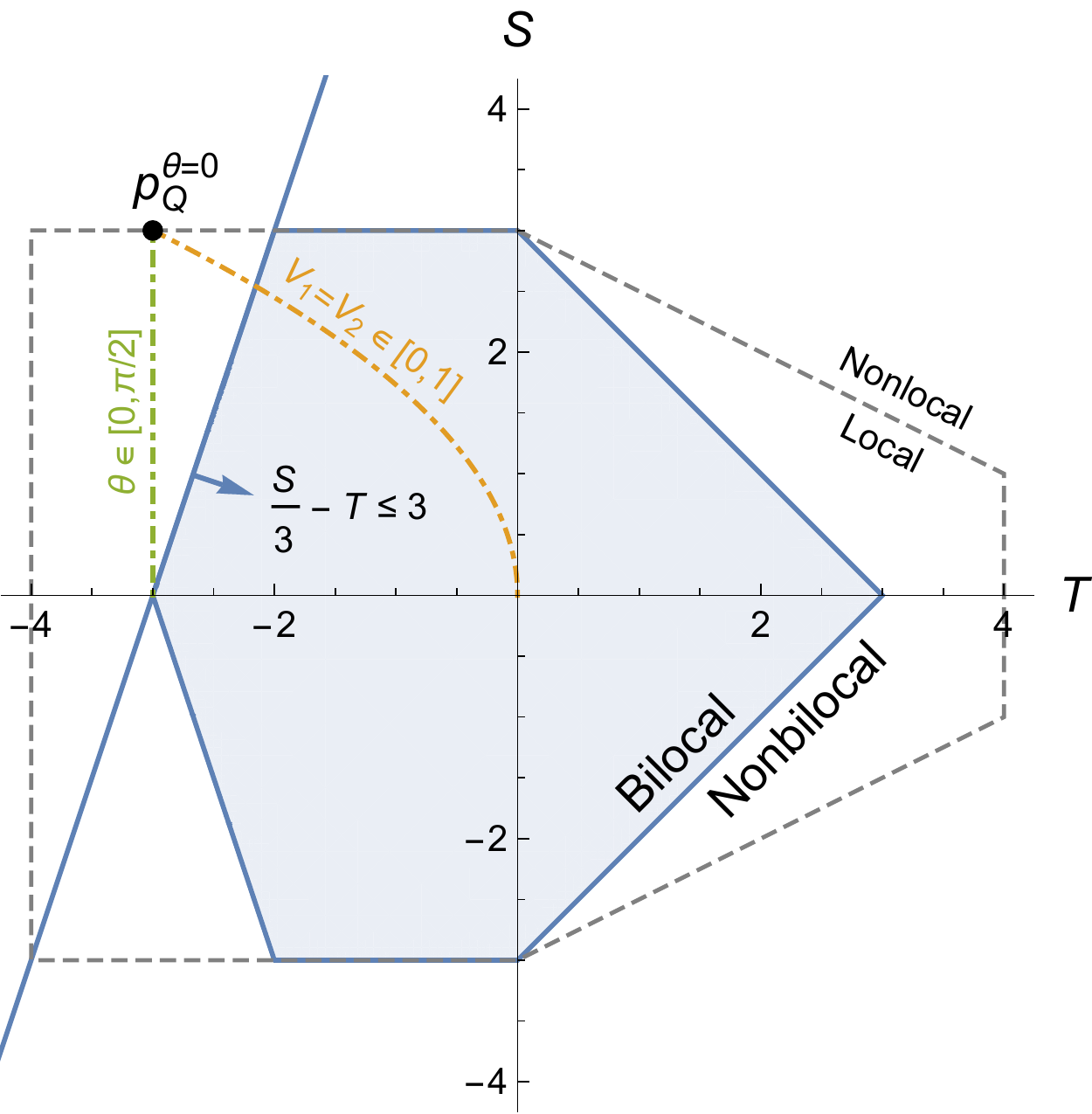}
	\caption{Projection of the ``$Z=0$ slice'' of the correlation space onto the $(S,T)$-plane. The blue region represents the projection of the bilocal set, delimited by the inequalities in Eq.~\eqref{all_biloc_ineq_in_slice}. The gray dashed lines delimit the projection of the local set, according to Eq.~\eqref{all_loc_ineq_in_slice}. The black point shows the projection of the quantum correlation $p_\text{Q}^{\theta=0}$ in the noiseless case ($V_1=V_2=0$). The orange and green dash-dotted curves show the projections of $p_\text{Q}^{\theta=0}$ for symmetric noise $V_1=V_2 \in [0,1]$ and the projections of $p_\text{Q}^{\theta}$ for all $\theta \in [0,\frac{\pi}{2}]$ in the noiseless case, respectively, with the former entering the bilocal set for visibilities $V_1=V_2=V_\text{crit}$ given by Eq.~\eqref{Vcrit_ineq}, and the latter remaining nonbilocal as long as $\theta < \frac{\pi}{2}$.}\label{ST_Projection_Z0_slice}
\end{figure}

To get some idea of what the set of bilocal correlations looks like, it is instructive to look at the \emph{projection} onto the $(S,T)$ plane of its \emph{slice} where $Z=0$. This projection, obtained through numerical optimisation to check the (non)bilocality of various points $(S,T)$, is shown on Figure~\ref{ST_Projection_Z0_slice}. Quite remarkably, and contrarily to all (nontrivial, multidimensional) bilocal sets previously studied in the literature, it appears that the bilocal set in this projected slice is delimited by linear inequalities, namely:
\begin{align}
\pm \frac{S}{3} - T \stackrel{\underset{Z=0}{\text{biloc}}}{\leq} 3, \quad \pm S \stackrel{\underset{Z=0}{\text{biloc}}}{\leq} 3, \quad \pm S + T \stackrel{\underset{Z=0}{\text{biloc}}}{\leq} 3. \label{all_biloc_ineq_in_slice}
\end{align}
We also verified these six inequalities via numerical optimisations, as we did for our other bilocal inequalities presented in this paper. These can also be proven for bilocal models with the tetrahedral symmetry in the same way as in the previous appendix.
The first of these inequalities, with a $+$ sign, corresponds precisely to our bilocal inequality~\eqref{bilocineq} for the $Z=0$ case. As we see, it thus appears to be ``tight'', in the sense of defining a facet of the bilocal set in the projected $Z=0$ slice.

To complete the picture, one may also look at the set of local correlations. This forms a convex polytope in the full correlation space, so it is expected to also be delimited by linear inequalities in the projected $Z=0$ slice. We find here that its facets are defined by
\begin{align}
\pm S \stackrel{\underset{Z=0}{\text{loc}}}{\leq} 3, \quad \pm T \stackrel{\underset{Z=0}{\text{loc}}}{\leq} 4, \quad \pm S + \frac{T}{2} \stackrel{\text{loc}}{\leq} 3 \label{all_loc_ineq_in_slice}
\end{align}
(with the last pair of inequalities holding in fact for general local models, without the $Z=0$ restriction); see Figure~\ref{ST_Projection_Z0_slice}. We note that the correlations $p_\text{Q}^\theta$ satisfy the above inequalities and, more generally, they admit a local model.

\section{$Z$-correction of the bilocal inequality}\label{AppCorrection}
As we just saw, when restricting to the case where $Z=0$, the bilocal inequality presented in Eq.~\eqref{bilocineq} of the main text is tight in the $(S,T)$ plane. However, when $Z$ is perturbed away from zero (e.g.~due to small experimental errors), then Eq.~\eqref{bilocineq} is not tight anymore.

\begin{figure}[h!]
	\centering
	\includegraphics[width=\columnwidth]{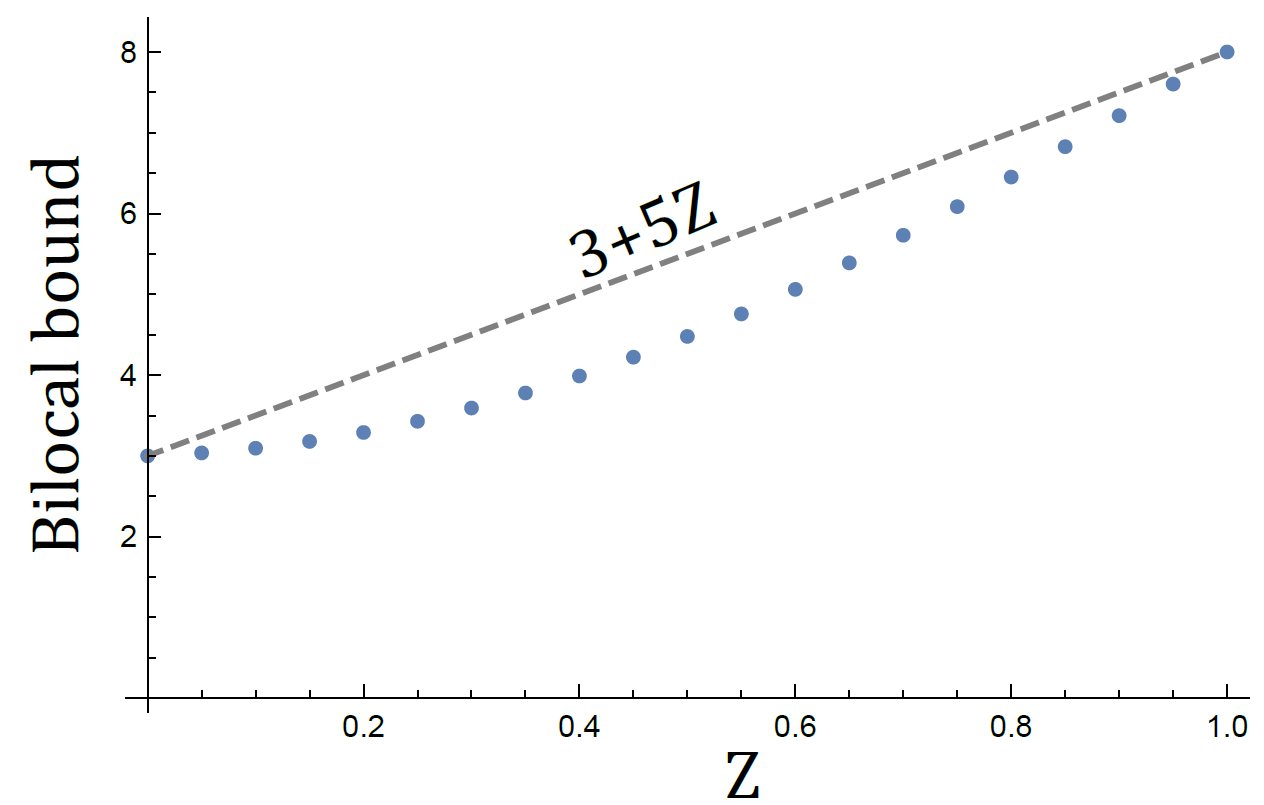}
	\caption{Results for the numerical optimisation of the bilocal bound of $\mathcal{B}$ for various values of $Z$ (blue dots), and a linear correction ($5Z$) to the bilocal bound of 3 associated to the case of $Z=0$.}\label{ZvsBiloc}
\end{figure}

We have numerically computed the largest values of $\mathcal{B} = \frac{S}{3}-T$ attainable for a given value of $Z$. This can be efficiently incorporated into the optimisation by placing the linear constraint $-Z\leq \langle \cdot \rangle \leq Z$ on all the one-, two- and three-party correlators that do not appear in $S$ and $T$. The results of the optimisation are displayed in Figure~\ref{ZvsBiloc}. The simplest correction term that can be added to the bilocal bound for $Z=0$ in order to account for the case where $Z>0$ is a linear correction of $5Z$, as illustrated and as we considered in Eq.~\eqref{bilocineq}. However, it is clear that more precise correction terms to the bilocal bound are also possible. 

\section{A second bilocal inequality}\label{AppSecondIneq}	

We detail the second  bilocal inequality mentioned in the main text. Alike the first bilocal inequality, it is inspired by the quantum correlations based on the EJM. In order to detect non-bilocal correlations without imposing additional constraints on the correlations, it is typically necessary to employ nonlinear expressions to capture the non-convexity of the set of bilocal correlations. We thus consider here the Bell expression (Eq.~\eqref{eq:SecondIneq} of the main text)
\begin{align}\nonumber\label{EJMineq}
\mathcal{B}' \equiv & \sum_{x, b}\sqrt{p(b)\left(1-b^x E^\text{A}_b(x)\right)}+\sum_{z, b}\sqrt{p(b)\left(1+b^z E^\text{C}_b(z)\right)} \\
& \quad + \sum_{x\neq z, b}\sqrt{p(b)\left(1-b^x b^z E^\text{AC}_b(x,z)\right)},
\end{align}
where we have defined the conditional one-party correlators $E^\text{A}_{b}(x)\equiv \sum_{a,c} a \, p(a,c|b,x,z)$ and $E^\text{C}_{b}(z)\equiv \sum_{a,c} c\, p(a,c|b,x,z)$ and the conditional two-party correlators $E^\text{AC}_{b}(x,z)\equiv \sum_{a,c} ac \, p(a,c|b,x,z)$, and where as in the main text Bob's output $b$ is written as $b = (b^1,b^2,b^3)$, with each $b^y = \pm 1$.
In terms of the (non-conditional) correlators considered previously, one has $p(b) = \frac14(1 + \sum_y b^y \langle B^y\rangle)$, $p(b)E^\text{A}_{b}(x) = \frac14(\langle A_x\rangle + \sum_y b^y \langle A_xB^y\rangle)$, $p(b)E^\text{C}_{b}(z) = \frac14(\langle C_z\rangle + \sum_y b^y \langle B^yC_z\rangle)$, and $p(b)E^\text{AC}_{b}(x,z) = \frac14(\langle A_xC_z\rangle + \sum_y b^y \langle A_xB^yC_z\rangle)$.

\subsection{Bilocal bound}

The bilocal bound on $\mathcal{B}'$ was obtained numerically, by optimising general models using two different numerical search methods~\cite{NumericalFootNote}. We found in particular that the same upper bound was obtained (up to machine precision) by bilocal models with the tetrahedral symmetry considered before; let us thus consider such models to obtain the analytical expression for the bilocal bound. 

Recall that for these models, $Z=0$---i.e., $\langle A^x\rangle = \langle B^y\rangle = \langle C^z\rangle = \langle A^x C^z\rangle = 0$; the bipartite correlators $\langle A_xB^y\rangle$ and $\langle B^yC_z\rangle$ are also $0$ whenever $x\neq y$ and $y\neq z$, resp.; and the tripartite correlators $\langle A_xB^yC_z\rangle$ are also $0$ whenever $x,y,z$ are not all different. It follows that $p(b)=\frac14$, $E^\text{A}_{b}(x) = b^x \langle A_xB^x\rangle$, $E^\text{C}_{b}(z) = b^z \langle B^zC_z\rangle$, and $E^\text{AC}_{b}(x,z) = \delta_{x\neq z} b^x b^z \langle A_xB^{y \neq x,z}C_z\rangle$ (where we used the fact that $b^1b^2b^3=1$, and where the superscript $y \neq x,z$ denotes the unique value of $y$ different from both $x$ and $z$ when $x\neq z$), so that $\mathcal{B}'$ can be written as
\begin{align}\nonumber
\mathcal{B}' & = 2\sum_x\sqrt{1-\langle A_xB^x\rangle}+2\sum_z\sqrt{1+\langle B^zC_z\rangle} \\
& \quad + 2\sum_{x\neq z}\sqrt{1-\langle A_xB^{y \neq x,z}C_z\rangle}.
\end{align}
Using Eq.~\eqref{correlators_sym_model} we obtain more specifically, in terms of the 14 parameters $q_+^{(1)}, q_+^{(2)}, q^{\tau_\alpha,\tau_\gamma}_{b=\tilde\alpha=\tilde\gamma|\tilde\alpha=\tilde\gamma}, q^{\tau_\alpha,\tau_\gamma}_{b=\tilde\alpha|\tilde\alpha\neq\tilde\gamma}, q^{\tau_\alpha,\tau_\gamma}_{b=\tilde\gamma|\tilde\alpha\neq\tilde\gamma}$ defining a symmetric bilocal model (see Appendix~\ref{AppSimulation}):
\begin{align}\nonumber
\mathcal{B}' & = 6\sqrt{\sum_{\tau_\alpha,\tau_\gamma} \! q_{\tau_\alpha}^{(1)}q_{\tau_\gamma}^{(2)} \Big(1 - \tau_\alpha \, q^{\tau_\alpha,\tau_\gamma}_{b=\tilde\alpha|\tilde\alpha\neq\tilde\gamma} + \tau_\alpha {\textstyle \frac{1-q^{\tau_\alpha,\tau_\gamma}_{b=\tilde\alpha=\tilde\gamma|\tilde\alpha=\tilde\gamma}}{3}}\Big)} \\
& \ \ + 6\sqrt{\sum_{\tau_\alpha,\tau_\gamma} \! q_{\tau_\alpha}^{(1)}q_{\tau_\gamma}^{(2)} \Big(1 + \tau_\gamma \, q^{\tau_\alpha,\tau_\gamma}_{b=\tilde\gamma|\tilde\alpha\neq\tilde\gamma} - \tau_\gamma {\textstyle \frac{1-q^{\tau_\alpha,\tau_\gamma}_{b=\tilde\alpha=\tilde\gamma|\tilde\alpha=\tilde\gamma}}{3}}\Big)} \nonumber \\
& \ \ + 12\sqrt{\sum_{\tau_\alpha,\tau_\gamma} \! q_{\tau_\alpha}^{(1)}q_{\tau_\gamma}^{(2)} \!\left(\!\!
\begin{array}{l}
1 - \tau_\alpha \tau_\gamma \frac12 + \tau_\alpha \tau_\gamma {\textstyle \frac{1-q^{\tau_\alpha,\tau_\gamma}_{b=\tilde\alpha=\tilde\gamma|\tilde\alpha=\tilde\gamma}}{3}} \\[1mm]
 \qquad + \tau_\alpha \tau_\gamma {\textstyle \frac{q^{\tau_\alpha,\tau_\gamma}_{b=\tilde\alpha|\tilde\alpha\neq\tilde\gamma} + q^{\tau_\alpha,\tau_\gamma}_{b=\tilde\gamma|\tilde\alpha\neq\tilde\gamma}}{2}}
\end{array}
\!\!\right)} .
\end{align}

One can then use the trivial (and all saturable) bounds $-q^{+,+}_{b=\tilde\alpha|\tilde\alpha\neq\tilde\gamma} \le 0$, $-q^{+,-}_{b=\tilde\alpha|\tilde\alpha\neq\tilde\gamma} \le 0$, $-{\textstyle \frac{1-q^{-,+}_{b=\tilde\alpha=\tilde\gamma|\tilde\alpha=\tilde\gamma}}{3}} \le 0$ and $q^{-,-}_{b=\tilde\alpha|\tilde\alpha\neq\tilde\gamma} \le 1$ under the first square root, $q^{+,+}_{b=\tilde\gamma|\tilde\alpha\neq\tilde\gamma} \le 1$, $-q^{+,-}_{b=\tilde\gamma|\tilde\alpha\neq\tilde\gamma} \le 0$, $-{\textstyle \frac{1-q^{-,+}_{b=\tilde\alpha=\tilde\gamma|\tilde\alpha=\tilde\gamma}}{3}} \le 0$ and $-q^{-,-}_{b=\tilde\gamma|\tilde\alpha\neq\tilde\gamma} \le 0$ under the second square root, and $q^{+,+}_{b=\tilde\alpha|\tilde\alpha\neq\tilde\gamma} + q^{+,+}_{b=\tilde\gamma|\tilde\alpha\neq\tilde\gamma} \le 1$, $-q^{+,-}_{b=\tilde\alpha|\tilde\alpha\neq\tilde\gamma} - q^{+,-}_{b=\tilde\gamma|\tilde\alpha\neq\tilde\gamma} \le 0$, $-{\textstyle \frac{1-q^{-,+}_{b=\tilde\alpha=\tilde\gamma|\tilde\alpha=\tilde\gamma}}{3}} \le 0$ and $q^{-,-}_{b=\tilde\alpha|\tilde\alpha\neq\tilde\gamma} + q^{-,-}_{b=\tilde\gamma|\tilde\alpha\neq\tilde\gamma} \le 1$ under the third square root, to upper-bound $\mathcal{B}'$ above by a (saturable) expression that does no longer contain the 7 different parameters involved here.
This leaves us with only 7 (out of the initial 14) free parameters to optimise for the symmetric models, at which point we resort to numerical means. We find in particular that the maximum of the $\mathcal{B}'$ expression is obtained by choosing $q_+^{(1)}=q_+^{(2)}=1$ or $q_+^{(1)}=q_+^{(2)}=0$. In the first case, we thus obtain
\begin{equation}
\mathcal{B}'\leq 18\sqrt{1+{\textstyle \frac{1-q^{+,+}_{b=\tilde\alpha=\tilde\gamma|\tilde\alpha=\tilde\gamma}}{3}}}+6\sqrt{2-{\textstyle \frac{1-q^{+,+}_{b=\tilde\alpha=\tilde\gamma|\tilde\alpha=\tilde\gamma}}{3}}},
\end{equation}
which reaches its maximum for $q^{+,+}_{b=\tilde\alpha=\tilde\gamma|\tilde\alpha=\tilde\gamma} = 0$. Thus, we find the bilocal bound to be
\begin{equation}\label{bilocVal}
\mathcal{B}' \stackrel{\text{biloc}}{\leq} 12\sqrt{3} + 2\sqrt{15}\approx 28.53.
\end{equation}

We reiterate that, although obtained explicitly here for bilocal models with the tetrahedral symmetry, this bound was verified numerically to hold for general bilocal models.

\subsection{Quantum violations}

The bilocal bound is violated by the  quantum correlations $p_\text{Q}^\theta$, based on the family of measurements generalising the EJM. Evaluating the conditional one- and two-party correlators, we obtain
 \begin{align}
& E_b^\text{A}(x)=-b^x\frac{V_1}{2}\cos\theta,  \quad  E_b^\text{C}(z)=b^z\frac{V_2}{2}\cos\theta, \nonumber \\
& E_b^\text{AC}(x,z)= \begin{cases}
-b^xb^z\frac{V_1V_2}{2}\left(1{+}\sin\theta\right) & \text{if } xz\in\{13,21,32\}\\
-b^xb^z\frac{V_1V_2}{2}\left(1{-}\sin\theta\right) & \text{if } xz\in\{12,23,31\}\\
\,0 & \text{otherwise}
\end{cases}\!.
\end{align}
Together with $p_\text{Q}^\theta(b)=\frac14$ for all $b$, this gives
\begin{align}\nonumber
\mathcal{B}' & = 6\sqrt{1+{\textstyle \frac{V_1}{2}}\cos\theta}+6\sqrt{1+{\textstyle \frac{V_2}{2}}\cos\theta} \\
& \quad + 6\sqrt{1+{\textstyle \frac{V_1V_2}{2}}\left(1{+}\sin\theta\right)} + 6\sqrt{1+{\textstyle \frac{V_1V_2}{2}}\left(1{-}\sin\theta\right)}.
\end{align}

In the noiseless case ($V_1=V_2=1$), this gives a violation of the bilocal inequality~\eqref{bilocVal} for all $\theta$ in the range $0 \le \theta \lesssim 0.254\,\pi$, with a maximal value of
\begin{equation}\label{BQ}
\mathcal{B}'=12\sqrt{6}\approx 29.39,
\end{equation}
obtained for $\theta = 0$. For $\theta = 0$ precisely, allowing now for symmetric noise, we find a violation for visibilities $V_1=V_2$ larger than the critical visibility $V_\text{crit} \approx 88.0\,\%$.

\subsection{Comparison between our two bilocal inequalities}

From the results above it seems that our second bilocal inequality, Eq.~\eqref{bilocVal}, is less powerful than our first one, Eq.~\eqref{bilocineq}, at detecting the non-bilocality of the quantum correlation $p_\text{Q}^\theta$: indeed it detects it only for a restricted range of $\theta$, and for larger visibilities. More generally, we find that any choice of parameters $(V_1,V_2,\theta)$ for which $p_\text{Q}^\theta$ violates our second inequality already violates our first inequality.

However, looking beyond the specific quantum correlation $p_\text{Q}^\theta$, one can find non-bilocal correlations that violate Eq.~\eqref{bilocVal} but not Eq.~\eqref{bilocineq}. An example is for instance given by the (local) correlation defined by
\begin{align}
& \langle A_x\rangle = \langle B^y\rangle = \langle C_z\rangle = \langle A_xC_z\rangle = 0, \nonumber \\[1mm]
& \langle A_xB^y\rangle= -\frac12 \delta_{x,y}, \quad \langle B^yC_z\rangle= \frac12 \delta_{y,z} , \nonumber \\[1mm]
& \langle A_xB^yC_z\rangle= -\frac13 \delta_{x\neq y\neq z} ,
\end{align}
which gives $(S,T,Z) = (3,-2,0)$ and thus satisfies Eq.~\eqref{bilocineq} (and in fact, satisfies all inequalities~\eqref{all_biloc_ineq_in_slice} that bound the $(S,T)$-projection of the bilocal set when $Z=0$, see Appendix~\ref{AppSlice}), while $\mathcal{B}' = 6\sqrt{6} + 8\sqrt{3} \approx 28.55 > 12\sqrt{3} + 2\sqrt{15} \approx 28.53$ violates Eq.~\eqref{bilocVal}.

It is clear that many different bilocal inequalities could be obtained by considering various types of nonlinear functions of the correlations, as we did here with $\mathcal{B}'$. Some may be found to be better-suited for certain correlations of interest, other than $p_\text{Q}^\theta$.

\subsection{Stronger-than-quantum nonlocality}

We have also used the Bell expression~\eqref{EJMineq} to detect stronger-than-quantum network nonlocality. This is achieved by deriving a quantum Bell inequality for the network, i.e.~a non-trivial bound on $\mathcal{B}'$ satisfied by all quantum models with two 	independent sources. The bound is established under the mild restriction that Bob has uniform outcomes, i.e.~$p(b)=\frac{1}{4}$ for all $b$. To this end, we consider the use of a simple concavity inequality to linearize $\mathcal{B}'$: for any $a_1,\ldots,a_n\geq 0$, it holds that
\begin{equation}
\sum_{i=1}^n \sqrt{a_i}\leq \sqrt{n \sum_{i=1}^n a_i},
\end{equation}
with equality if and only if all $a_i$ are equal. Since the Bell expression $\mathcal{B}'$ is a sum of square-root expressions, using the concavity inequality above allows us to bound it with an expression that is a square-root of the corresponding sums. One thus finds
\begin{equation}
\mathcal{B}'\leq \sqrt{48\, \mathcal{B}_\text{lin.}'},
\end{equation}
where we have defined the linear expression	
\begin{align}\nonumber
\mathcal{B}_\text{lin.}' \equiv & \sum_{x, b} \frac14 \left(1-b^x E^\text{A}_b(x)\right) +\sum_{z, b} \frac14 \left(1+b^z E^\text{C}_b(z)\right) \\
& \quad + \sum_{x\neq z, b} \frac14 \left(1-b^x b^z E^\text{AC}_b(x,z)\right).
\end{align}

We can now bound $\mathcal{B}_\text{lin.}'$ for quantum models, with independent sources, by using the semidefinite relaxations of Ref.~\cite{Pozas}. Thanks to codes provided by A.~Pozas-Kerstjens, we have been able to evaluate the third level SDP relaxation described in~\cite{Pozas} and obtain $ \mathcal{B}_\text{lin.}' \lesssim 19.64$. This corresponds to $\mathcal{B}'\leq \sqrt{48\, \mathcal{B}_\text{lin.}'} \lesssim 30.70$.


\begin{thebibliography}{99}
		
\bibitem{Bell}
J. S. Bell,
On the Einstein Podolsky Rosen Paradox,
Physics \textbf{1}, 195 (1964).
	
\bibitem{Review}
N. Brunner, D. Cavalcanti, S. Pironio, V. Scarani, and S. Wehner,
Bell nonlocality,
\href{https://doi.org/10.1103/RevModPhys.86.419}{Rev. Mod. Phys. \textbf{86}, 419 (2014).}
		
\bibitem{GisinQchance14} N. Gisin, {\it Quantum Chance, nonlocality, teleportation and other quantum marvels}, Springer, 2014.
		
\bibitem{Branciard}
C. Branciard, N. Gisin, and S. Pironio,
Characterizing the Nonlocal Correlations Created via Entanglement Swapping,
\href{https://doi.org/10.1103/PhysRevLett.104.170401}{Phys. Rev. Lett. \textbf{104}, 170401 (2010).}
	
\bibitem{Fritz}
T. Fritz,
Beyond Bell's theorem: correlation scenarios,
\href{https://doi.org/10.1088/1367-2630/14/10/103001}{New J. Phys. \textbf{14} 103001 (2012).}
		
\bibitem{Zukowski}
M. \.Zukowski, A. Zeilinger, M. A. Horne, and A. K. Ekert,
``Event-ready-detectors`` Bell experiment via entanglement swapping,
\href{https://doi.org/10.1103/PhysRevLett.71.4287}{Phys. Rev. Lett. \textbf{71}, 4287 (1993).}
		
\bibitem{Branciard2}
C. Branciard, D. Rosset, N. Gisin, and S. Pironio,
Bilocal versus nonbilocal correlations in entanglement-swapping experiments,
\href{https://doi.org/10.1103/PhysRevA.85.032119}{Phys. Rev. A \textbf{85}, 032119 (2012).}
	
\bibitem{ChavesFritz}
R. Chaves and T. Fritz,
Entropic approach to local realism and noncontextuality,
\href{https://doi.org/10.1103/PhysRevA.85.032113}{Phys. Rev. A \textbf{85}, 032113 (2012).}
	
\bibitem{Tavakoli}
A. Tavakoli, P. Skrzypczyk, D. Cavalcanti, and A. Ac\'in,
Nonlocal correlations in the star-network configuration,
\href{https://doi.org/10.1103/PhysRevA.90.062109}{Phys. Rev. A \textbf{90}, 062109 (2014).}
	
\bibitem{Henson}
J. Henson, R. Lal and M. F. Pusey,
Theory-independent limits on correlations from generalised Bayesian networks,
\href{https://doi.org/10.1088/1367-2630/16/11/113043}{	New J. Phys. \textbf{16}, 113043 (2014).}
	
\bibitem{Wood}
C. J. Wood and R. W. Spekkens,
The lesson of causal discovery algorithms for quantum correlations: causal explanations of Bell-inequality violations require fine-tuning,
\href{https://doi.org/10.1088/1367-2630/17/3/033002}{New J. Phys. \textbf{17} 033002 (2015).}

\bibitem{ChavesKueng}
R. Chaves, R. Kueng, J. B. Brask, and D. Gross,
Unifying Framework for Relaxations of the Causal Assumptions in Bell's Theorem,
\href{https://doi.org/10.1103/PhysRevLett.114.140403}{Phys. Rev. Lett. \textbf{114}, 140403 (2015).}

\bibitem{TavakoliConnected}
A. Tavakoli,
Quantum Correlations in Connected Multipartite Bell Experiments,
\href{https://doi.org/10.1088/1751-8113/49/14/145304}{	J. Phys. A: Math and Theor \textbf{49}, 145304 (2016).}

\bibitem{Fritz2}
T. Fritz,
Beyond Bell's Theorem II: Scenarios with arbitrary causal structure,
\href{https://doi.org/10.1007/s00220-015-2495-5}{Comm. Math. Phys. \textbf{341}, 391-434 (2016).}		
	
\bibitem{Rosset}
D. Rosset, C. Branciard, T. J. Barnea, G. P\"utz, N. Brunner, and N. Gisin,
Nonlinear Bell Inequalities Tailored for Quantum Networks,
\href{https://doi.org/10.1103/PhysRevLett.116.010403}{Phys. Rev. Lett. \textbf{116}, 010403 (2016).}
	
\bibitem{Chaves}
R. Chaves,
Polynomial Bell Inequalities,
\href{https://doi.org/10.1103/PhysRevLett.116.010402}{Phys. Rev. Lett. \textbf{116}, 010402 (2016).}
	
\bibitem{Tavakoli3}
A. Tavakoli,
Bell-type inequalities for arbitrary noncyclic networks,
\href{https://doi.org/10.1103/PhysRevA.93.030101}{Phys. Rev. A \textbf{93}, 030101(R) (2016).}

\bibitem{Tavakoli2}
A. Tavakoli, M-O. Renou, N. Gisin, and N. Brunner,
Correlations in star networks: from Bell inequalities to network inequalities,
\href{https://doi.org/10.1088/1367-2630/aa7673}{New J. Phys. \textbf{19}, 073003 (2017).}
	
\bibitem{Andreoli}
F. Andreoli, G. Carvacho, L. Santodonato, R. Chaves and F. Sciarrino,
Maximal violation of n-locality inequalities in a star-shaped quantum network,
\href{https://doi.org/10.1088/1367-2630/aa8b9b}{New J. Phys. \textbf{19}, 113020 (2017).}
	
\bibitem{Fraser}
T. C. Fraser and E. Wolfe,
Causal compatibility inequalities admitting quantum violations in the triangle structure,
\href{https://doi.org/10.1103/PhysRevA.98.022113}{Phys. Rev. A \textbf{98}, 022113 (2018).}
	
\bibitem{Luo}
M-X. Luo,
Computationally Efficient Nonlinear Bell Inequalities for Quantum Networks,
\href{https://doi.org/10.1103/PhysRevLett.120.140402}{Phys. Rev. Lett. \textbf{120}, 140402 (2018).}
	
\bibitem{Inflation}
E. Wolfe, R. W. Spekkens, and T. Fritz,
The Inflation Technique for Causal Inference with Latent Variables,
\href{https://doi.org/10.1515/jci-2017-0020}{J. Causal Inference \textbf{7}, 2 (2019).}

\bibitem{Wolfe}
E. Wolfe, A. Pozas-Kerstjens, M. Grinberg, D. Rosset, A. Ac\'in, and M. Navascues,
Quantum Inflation: A General Approach to Quantum Causal Compatibility,
\href{https://arxiv.org/abs/1909.10519}{arXiv:1909.10519}

\bibitem{Salman}
M-O. Renou, E. B\"aumer, S. Boreiri, N. Brunner, N. Gisin, and S. Beigi,
Genuine Quantum Nonlocality in the Triangle Network,
\href{https://doi.org/10.1103/PhysRevLett.123.140401}{Phys. Rev. Lett. \textbf{123}, 140401 (2019).}	

\bibitem{Renou}
M-O. Renou, Y. Wang, S. Boreiri, S. Beigi, N. Gisin, and N. Brunner,
Limits on Correlations in Networks for Quantum and No-Signaling Resources,
\href{https://doi.org/10.1103/PhysRevLett.123.070403}{Phys. Rev. Lett. 123, 070403 (2019).}

\bibitem{GisinBancal}
N. Gisin, J-D. Bancal, Y. Cai, A. Tavakoli, E. Z. Cruzeiro, S. Popescu, and N. Brunner,
Constraints on nonlocality in networks from no-signaling and independence,
\href{https://doi.org/10.1038/s41467-020-16137-4}{Nat Commun \textbf{11}, 2378 (2020)}.

\bibitem{Teleport}
C. H. Bennett, G. Brassard, C. Cr\'epeau, R. Jozsa, A. Peres, and W. K. Wootters,
Teleporting an unknown quantum state via dual classical and Einstein-Podolsky-Rosen channels,
\href{https://doi.org/10.1103/PhysRevLett.70.1895}{Phys. Rev. Lett. \textbf{70}, 1895 (1993).}
	
\bibitem{Gisin}
N. Gisin, Q. Mei, A. Tavakoli, M-O. Renou, and N. Brunner,
All entangled pure quantum states violate the bilocality inequality,
\href{https://doi.org/10.1103/PhysRevA.96.020304}{Phys. Rev. A \textbf{96}, 020304(R)  (2017).}
		
\bibitem{CHSH} J. F. Clauser, M. A. Horne, A. Shimony, and R. A. Holt,
Proposed Experiment to Test Local Hidden-Variable Theories,
\href{https://doi.org/10.1103/PhysRevLett.23.880}{Phys. Rev. Lett. \textbf{23}, 880 (1969).}
	
\bibitem{EJM}
N. Gisin,
Entanglement 25 Years after Quantum Teleportation: Testing Joint Measurements in Quantum Networks,
\href{https://doi.org/10.3390/e21030325}{Entropy \textbf{21}, 325 (2019).}
	
\bibitem{werner} R. F. Werner,
Quantum states with Einstein-Podolsky-Rosen correlations admitting a hidden-variable model,
\href{https://doi.org/10.1103/PhysRevA.40.4277}{Phys. Rev. A \textbf{40}, 4277 (1989).}


\bibitem{CommentNote}
Another natural scenario is to let Alice and Charlie perform four measurements, with Bloch vectors pointing to the vertices of a tetrahedron. It turns out, however, that the resulting correlations are less robust to noise than those obtained from measuring the three Pauli observables.
	
\bibitem{footnote_p_from_correlators} Specifically, one has $p(a,b^1,b^2,b^3,c|x,z) = \frac{1}{16}\big[1 + a \langle A_x\rangle + \sum_y b^y \langle B^y\rangle + c \langle C_z\rangle + \sum_y a b^y \langle A_xB^y\rangle + \sum_y b^y c \langle B^yC_z\rangle + a c \langle A_xC_z\rangle + \sum_y a b^y c \langle A_xB^yC_z\rangle \big]$. Notice the (convenient) redundancy in our encoding of Bob's outcome, as the product of its three $\pm1$-valued bits $b^1 b^2 b^3$ is always $+1$. As in Ref.~\cite{Branciard2}, we write $y$ as superscripts in $b^y, B^y$ to distinguish the case where the outputs ($b^y$) are all observed together, from the case of outputs obtained for different inputs (as in $A_x, C_z$).
	
\bibitem{Pozas}
A. Pozas-Kerstjens, R. Rabelo, L. Rudnicki, R. Chaves, D. Cavalcanti, M. Navascu\'es, and A. Ac\'in,
Bounding the Sets of Classical and Quantum Correlations in Networks,
\href{https://doi.org/10.1103/PhysRevLett.123.140503}{Phys. Rev. Lett. \textbf{123}, 140503 (2019).}
	
\bibitem{Pozas2}
Private communication with A. Pozas-Kerstjens.	

\bibitem{NumericalFootNote}
We have used the search method based on the Fourier transform of $p_\text{biloc}$ and standard brute-force search using Matlab's fmincon module to confirm our bilocal inequalities.

\bibitem{Lutkenhaus}		
N. L\"utkenhaus, J. Calsamiglia, and K-A. Suominen,
On Bell measurements for teleportation,
\href{https://doi.org/10.1103/PhysRevA.59.3295}{Phys. Rev. A \textbf{59}, 3295 (1999).}

\bibitem{Loock}
P. van Loock and N. L\"utkenhaus,
Simple criteria for the implementation of projective measurements with linear optics,
\href{https://doi.org/10.1103/PhysRevA.69.012302}{Phys. Rev. A \textbf{69}, 012302 (2004).}		
	
\bibitem{Barenco}
A. Barenco, C. H. Bennett, R. Cleve, D. P. DiVincenzo, N. Margolus, P. Shor, T. Sleator, J. A. Smolin, and H. Weinfurter,
Elementary gates for quantum computation,
\href{https://doi.org/10.1103/PhysRevA.52.3457}{Phys. Rev. A \textbf{52}, 3457 (1995).}

\bibitem{Nielsen}
M. A. Nielsen  and I. L. Chuang,
Quantum Computation and Quantum Information (10th Anniversary edition)
Cambridge University Press, 2010.

\end{thebibliography}
\end{document}